\documentclass[aps,twocolumn]{revtex4}
\usepackage{graphicx}   
\usepackage{latexsym}   

\newcommand{\bld}[1]{\mbox{\boldmath $#1$}}     
\newcommand{\R}{\bld{R}}	     		
\newcommand{\etal}{{\em et al.}}		
\newcommand{\ie}{{\em i.e.}}			
\newcommand{\MeV}{{\rm MeV}}			

\newcommand{\beq}{\begin{equation}}
\newcommand{\eeq}{\end{equation}}
\newcommand{\beqar}{\begin{eqnarray}}
\newcommand{\eeqar}{\end{eqnarray}}
\newcommand{\quart}{\mbox{{$1\over4$}}}
\newcommand{\half}{\mbox{{$1\over2$}}}
\newcommand{\threehalf}{\mbox{{$3\over2$}}}

\newcommand{\del}{\partial}

\newcommand{\s}{\sigma}		

\newcommand{\SA}{\bld{S}_A} \newcommand{\sA}{\bld{s}_A}
\newcommand{\SB}{\bld{S}_B} \newcommand{\sB}{\bld{s}_B}
\newcommand{\SF}{\bld{S}_F}

\newcommand{\IA}{{\cal I}_A}
\newcommand{\IB}{{\cal I}_B}

\newcommand{\IR}{{\cal I}_R}

\newcommand{\IF}{{\cal I}_F}

\newcommand{\Itot}{{\cal I}_0}

\newcommand{\Iwrig}{{\cal I}_{\rm wrig}}
\newcommand{\Ibend}{{\cal I}_{\rm bend}}
\newcommand{\Itwst}{{\cal I}_{\rm twst}}

\newcommand{\It}{{\cal I}_\perp}
\newcommand{\Iz}{{\cal I}_\parallel}

\newcommand{\para}{\parallel}
\newcommand{\zs}{{\rm zs}}

\begin{document}

\title{Correlated fission fragment spin dynamics}
\author{J\o rgen Randrup$^1$, Pavel Nadtochy$^2$,
Christelle Schmitt$^{3,4}$, and Katarzyna Mazurek$^4$}

\affiliation{
  $^1$Nuclear Science Division,  Lawrence Berkeley National Laboratory,
  Berkeley, CA 94720, USA \break
  $^2$Instituto Nazionale di Fisica Nucleare, Sezione di Napoli, 
  	Naples, Italy \break
  $^3$Institut Pluridisciplinaire Hubert Curien, Strasbourg, France \break
  $^4$Institute of Nuclear Physics, 
       Polish Academy of Sciences, Krakov, Poland}

\date{March 29, 2026}

\begin{abstract}
This study explores the role of nucleon exchange
for the generation of the fission fragment angular momenta.
For a number of typical fission cases, samples of $10^4$ shape evolutions
are generated by Langevin simulation and, subsequently, for each such evolution,
the nucleon exchange transport theory previously developed for damped nuclear
reactions is used to obtain the development of the fragment spin-spin 
distribution within the Fokker-Planck transport framework.
The characteristic evolution of both parallel and perpendicular 
spin components is discussed. A common feature is that the rotational modes
fall out of equilibrium before scission when the temperature rises rapidly
while the concurent shrinking of the neck suppresses further exchange.
A number of fission observables are extracted from the event ensembles:
the distribution of the magnitude of the fragment spin
and its orientation relative to the fission axis,
as well as the correlation between the two spins
and the distribution of their opening angle.
The dependence of these observables on the mass asymmetry is also examined.
\end{abstract}

\maketitle

\section{Introduction}

In low-energy nuclear fission, the two primary fragments emerge
with typically half a dozen units of angular momentum,
and more at higher energies.
But the underlying mehanism generating the angular momenta 
is not yet understood and the topic is currently a very active research area.
A variety of mechanisms have been advocated 
\cite{{MorettoPRC21},
BertschPRC99,
VogtPRC103,
BulgacPRL126,
MarevicPRC104,
RandrupPRL127,
BulgacPRL128,
RDV,
RandrupPRCL106,
RandrupPRCL108,
DossingPRC109,
ScampsPRC108,
ScampsPRC110,
ShneidemanPRC111,
Nature590}
and many recent correlation experiments have illuminated the issue 
\cite{Nature590,
AdiliEPJ256,
TravarPLB817,
MarinPRC104,
StetcuPRL127,
GihaPRC107,
MarinPRC105,
GjestvangPRC108,
GihaPRC111}.

This study explores how nucleon exchange between the fledging fission fragments
affects the evolution of their angular momenta.
A transport theory for this mechanism was originally developed
for damped nuclear reactions \cite{NPA327,NPA383}
and it was found that multiple transfers of individual nucleons
can account for the large loss of kinetic energy,
the diffusive evolution of the mass and charge partition between the 
reaction partners, and the build up of their angular momenta \cite{SH}.

In fission, as the system evolves from the saddle region towards scission
it develops an ever more binary geometry,
resembling those characteristic of damped reactions.
It may therefore be expected that nucleon exchange will play a role
at the late stage of fission.

A detailed study of the effect of nucleon exchange on the 
dynamical evolution of the coupled spins in a binary system \cite{NPA433}
derived expressions for the mobility coefficients governing the
evolution of the dinuclear modes of rotation and the associated relaxation times.
The present investigation employs this formal framework 
for the evolution of the fragment spin distribution
as the fissioning system evolves from the saddle region to scission.

The investigation is carried out as follows.
In Sect.\ \ref{shapes}, ensembles of shape evolutions are generated
by means of an established Langevin simulation code,
for a number of typical fission cases.
For each such evolution, the time-dependent spin-spin distribution 
is obtained by means of the nucleon-exchange transport theory.
The correlated fragment spins and the associated normal modes of rotation
are reviewed in Sect.\ \ref{spins}
and their transport treatment is described in Sect.\ \ref{FP}.
In Sect.\ \ref{events},
the spin dynamics is discussed for a few illustrative events,
both for the components parallel to the fission axis
and for the more complicated perpendicular components.
Subsequently, in Sect.\ \ref{ensembles}, 
for the various fission cases considered,
the event ensembles are analyzed and 
results for the fragment spin distributions are presented.

\section{Shape dynamics}
\label{shapes}

The dynamical evolution of the fissioning nucleus
is simulated with a well established Langevin treatment \cite{Adeev2005}.
It describes the shape of the nuclear system by means of three parameters
introduced in Ref.\ \cite{BrackRMP44}, equivalent to
the overall elongation, $R$, the radius of the neck between the two parts, $c$,
and the degree of reflection asymmetry, $\alpha$.
The Langevin simulation is briefly described in the Appendix.

The three-dimensional surface expressing the shape-dependent
potential energy is calculated in the macroscopic limit
where shell and pairing modifications are ignored,
as is apppropriate at high excitation.
Furthermore, the inertial-mass tensor associated with the shape motion
corresponds to incompressible irrotational flow, 
as has long been the standard choice.
Finally, the coupling between the shape parameters
and the remaining degrees of freedom is taken as the one-body dissipation,
expressed by means of the wall and window formulas \cite{onebody}.

The system is started at its ground-state shape
with a specified excitation energy.
It is then propagated with the Langevin equation 
which produces a Brownian-like evolution of the shape parameters. 
The system is followed as it explores the potential-energy surface
and eventually moves across the fission barrier region,
after which the neck radius quickly shrinks and scission occurs.

It is assumed that the total angular momentum of the system is zero.
This simplification is justified by the fact that,
in the excitation energy regime considered,
the contributions to the fragment spins from the overall rotational motion
of the complex are overwhelmed by the spin fluctuations
so our conclusions would not be affected.

In the present study, we consider fission of $^{182}$Hg, $^{202}$Po,
and $^{236}$U at the excitation energy $E^*=46\,\MeV$,
and in the latter case we also consider $E^*=70\,\MeV$ in order to explore
the energy dependence of the results.
For each of these four systems, two different values of the wall dissipation
strength are used, namely the originally derived strength ($k_s=1$) 
\cite{onebody} and a quarter of that ($k_s=0.25$)  \cite{Nix1986,Nix1987}.
which yields better agreement with various experimental data.

\begin{figure}[tbh]	    
\includegraphics[trim={0.2cm 0.4cm 0.2cm 0.4cm},clip,width=0.2\textwidth,
     angle=+90]{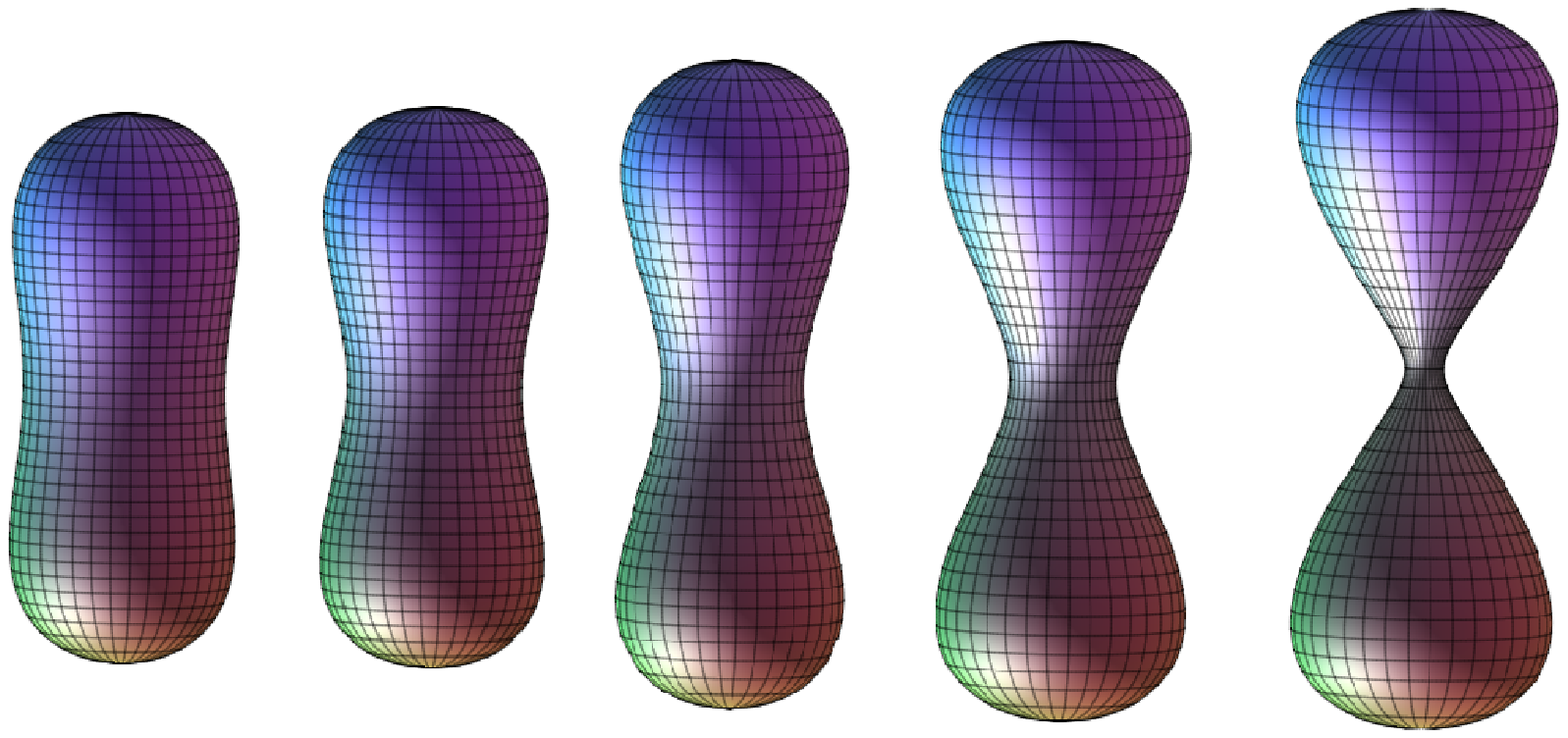}      
\caption{\label{f:shapes}
Five snapshots of the shape of $^{236}$U($E^*$=46\,MeV)
calculated with $k_s=1$, as it evolves from saddle to scission
for an event leading to a symmetic division.}
\end{figure}

For each of these eight cases,
the Langevin treatment is used to generate a sample of $10^4$ shape evolutions.
One of these is depicted in Fig.\ \ref{f:shapes}.
Subsequently, for each of the shape evolutions,
the time-dependent correlated spin distribution is obtained 
within the Fokker-Planck transport framework (see Sect.\ \ref{FP}).

\begin{figure}[tbh]	    
\includegraphics[trim={2.5cm 0.2cm 8cm 8.0cm},clip,width=0.28\textwidth,
     angle=-90]{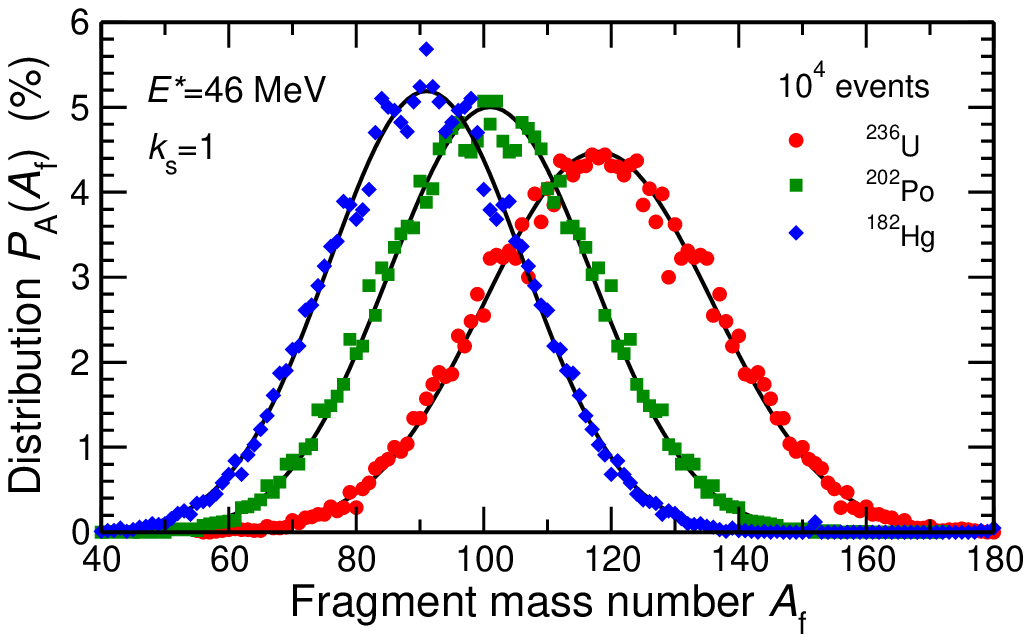}      
\caption{\label{f:Af} The fragment mass distributions from fission of
$^{236}$U, $^{202}$Po, and $^{182}$Hg at $E^*=46\,\MeV$
as generated by Langevin simulations of $10^4$ events (solid symbols), 
together with the associated Gaussians (solid curves).}
\end{figure}

Because the potential energy is calculated in the macroscopic limit,
the resulting fragment mass distributon, $P(A_{\rm f})$, 
is a single peak centered at symmetry,
and in each case its shape is very well represented by a gaussian function 
having the same width as the distribution, 
as illustrated in Fig.\ \ref{f:Af}.
It should be noted that even though the mass distributions are symmetric,
their large widths show that many of the events have large asymmetries,
making it practical to gate on the mass asymmetry.

\begin{figure}[tbh]	    
\includegraphics[trim={1.6cm 0cm 1cm 14cm},clip,width=0.52\textwidth,
     angle=-90]{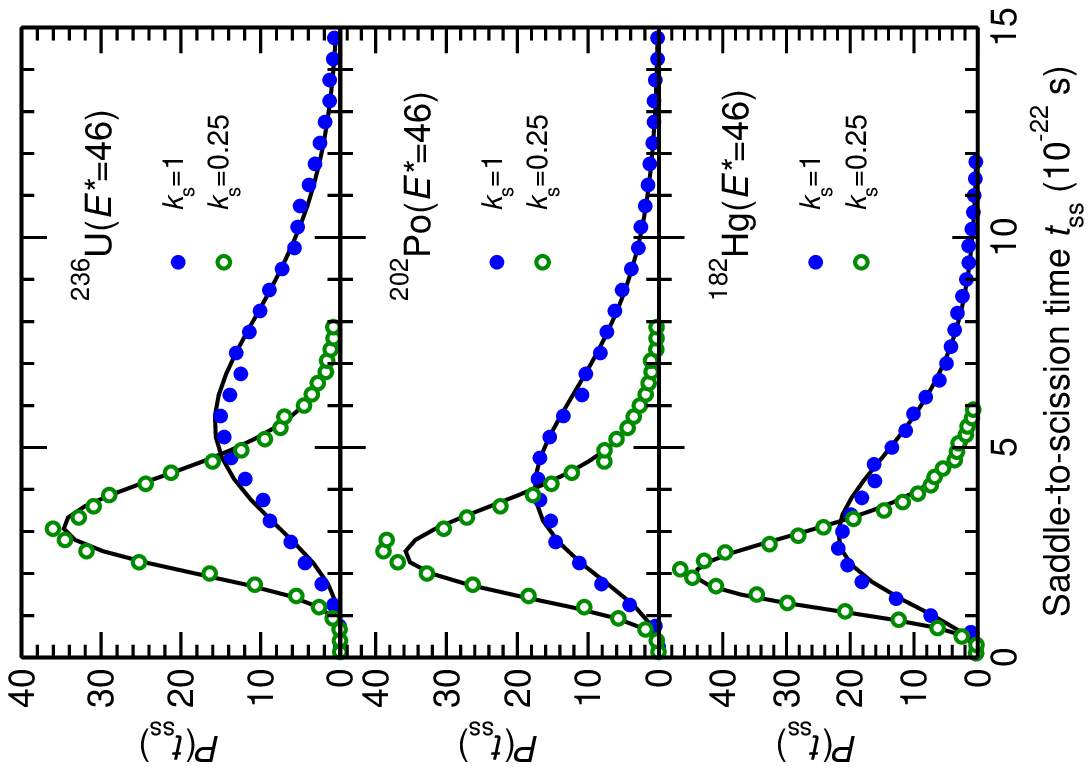}      
\caption{  \label{f:tss}
The distributions of the saddle-to-scission times $t_{\rm ss}$
for the six cases having $E^*$=46\,MeV, 
as obtained with the Langevin simulations of the shape evolution (symbols), 
and the associated gamma distributions (solid curves).}
\end{figure}

The time spent by the system evolving from saddle to scission, $t_{\rm ss}$,
represents the time available for the spin transport process
and is a key quantity for the outcome of the spin evolution.
Generally, $t_{\rm ss}$ fluctuates considerably from one event to another
due to the stochastic character of the shape evolution.
The associated duration distribution, $P_{\rm ss}(t_{\rm ss})$,
is shown in Fig.\ \ref{f:tss} for various cases.
It is noteworthy that, for all the cases considered,
$P_{\rm ss}(t_{\rm ss})$ is very well represented by a gamma distribution,
\beq
P_{\rm ss}(t_{\rm ss})\ \approx\ 
       {1\over\theta_{\rm ss}}{1\over\Gamma(\alpha_{\rm ss})}
       \left({t_{\rm sss}\over\theta_{\rm ss}}\right)^{\alpha_{\rm ss}-1}
       e^{-t_{\rm ss}/\theta_{\rm ss}}\ .
\eeq
The parameters are obtained by matching the average duration $\bar{t}_{\rm ss}$ 
and the associated variance $\sigma_{\rm ss}^2$.
This yields the scale parameter as
$\theta_{\rm ss}=\sigma_{\rm ss}^2/\bar{t}_{\rm ss}$,
while the shape parameter is 
$\alpha_{\rm ss}=\bar{t}_{\rm sss}^2/\sigma_{\rm ss}^2-1$.
These representations of the duration distributions
are also shown in Fig.\ \ref{f:tss}.

As will be discussed in Sect.\ \ref{FP},
the quantities needed for the spin transport treatment are the shape parameters
$R(t)$, $c(t)$, and $\alpha(t)$, as well as the temperature $T(t)$.
Because the Langevin calculation employs time steps that are much shorter 
than the time scales characteristic of the spin evolution,
the very rapidly fluctuating Langevin results
are replaced by smooth functions for each shape evolution,
as illustrated in the Appendix.
This simplification is made for practical connvenience
and has no influence on the calculated spin evolution.

\begin{figure}[tbh]	    
\includegraphics[trim={2.5cm 0.1cm 1.5cm 8.0cm},clip,width=0.4\textwidth,
     angle=-90]{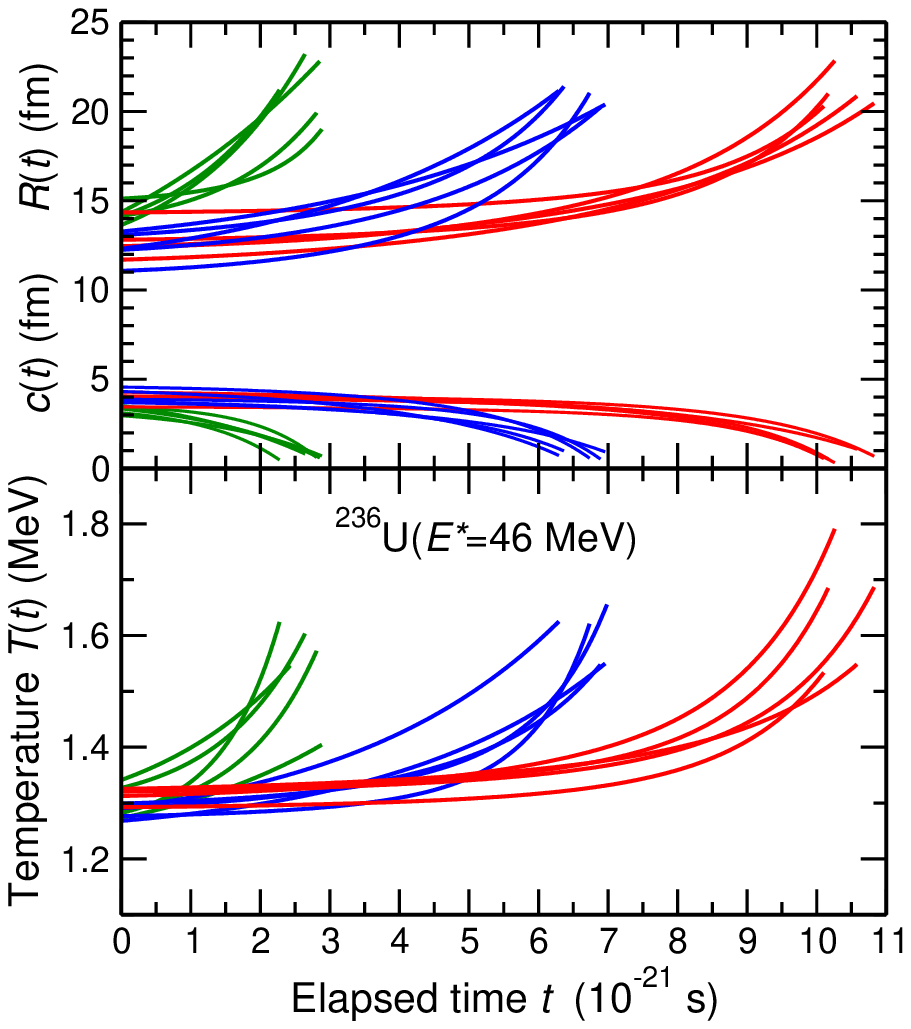}      
\caption{\label{f:RcT}
The smoothed time evolution of the elongation $R$,
the neck radius $c$, and the temperature $T$,
for five Langevin shape evolutions of
short (green), medium (blue), or long (red) duration,
for fission of $^{236}$U at $E^*=46\,\MeV$ using $k_s=1$.}
\end{figure}

The resulting smooth time evolutions of $R(t)$, $c(t)$, and $T(t)$
are shown in Fig.\ \ref{f:RcT} for events of various duration,
either ``fast'' ($2\,\zs\leq t_{\rm ss}\leq3\,\zs$),
``medium'' ($6\,\zs\leq t_{\rm ss}\leq7\,\zs$),
or ``slow'' (having $10\,\zs\leq t_{\rm ss}\leq11\,\zs$).
Because the driving force is small near the saddle,
the early part of the shape evolution is diffusion-dominated,
with the system never straying far from the saddle region.
By contrast, the disruptive driving force is large near scission,
so the late evolution is drift-dominated,
with the elongation and the temperature increasing rapily
as the neck is closing.
While the slow events spend most of the time in the diffusion-dominated region,
the fast events enter the drift-dominated region very quickly.
Generally,
the last evolution stage does not depend on the total duration of the event.

\section{Correlated fragment spins}
\label{spins}

As the fissioning nuclear system evolves from the saddle region
towards scission, it attains an ever more binary character.
The present study considers the system as a dinucleus during that
entire evolution.
While this idealization may be questionable early on
when the shape is still fairly compact, 
the treatment receives some justification a posteriori from the fact that
the calculated results are not very sensitive to the early evolution
due to the relatively short relaxation times for the spin transport.

In a dinuclear system, the transfer of individual nucleons 
between the two binary partners changes their linear and angular momenta
as well as their excitation energies.
This mechanism was found to be the dominant dissipation mechanism
in damped nuclear reactions \cite{SH} and we wish to here explore
its influence on the angular momenta of the emerging fission fragments.

The fledging fragments are considered as separate entities
whose individual angular momenta are denoted by $\SA$ and $\SB$,
and $\bld{L}=\R\times\bld{P}$ is the angular momentum 
associated with their relative motion,
where $\R\equiv\R_A-\R_B$ is the relative position of the two fragments
and $\bld{P}=\mu\dot{\bld V}$ is their relative momentum.
While \bld{L}\ is necessarily orthogonal to \R,
the individual spins $\SF$ may have components 
both perpendicular and parallel to \R.
The associated moments of inertia generally differ, $\IF^\perp\neq\IF^\para$,
and they are taken to be the rigid-body values,
as is appropriate at the temperature range considered.
(For simplicity, 
${\cal I}_F^\perp={\cal I}_F^\para=\mbox{$2\over5$}M_F R_F^2$ is used here.)

The rotational energy of the system can then be written as a sum,
$E_{\rm rot}=E_\para+E_\perp$, where
\beqar
E_\para &=& (S_A^\para)^2/2\IA^\para + (S_B^\para)^2/2\IB^\para\ ,\\
E_\perp &=& (S_A^\perp)^2/2\IA^\perp + (S_B^\perp)^2/2\IB^\perp + L^2/2\IR\ ,
\eeqar
with $\IR=\mu R^2$.
Because the system is isolated, the total angular momentum,
$\bld{J}=\SA+\SB+\bld{L}$, is constant, 
so $\bld{L}$ can be eliminated from $E_\perp$

The lowest rotational energy is obtained when the system rotates rigidly,
in which case the individual fragment spins are
$S_F^\parallel=(\IF^\para/\Iz)J_\parallel$,
where $\Iz=\IA^\parallel+\IB^\parallel$,
and $\SF^\perp=(\IF^\perp/\It)\bld{J}_\perp$,
where $\It=\IA^\perp+\IB^\perp+\IR$.
The energy associated with such a rigid rotation is
$E_{\rm rigid}=J_\para^2/2\Iz + J_\perp^2/2\It$.

It simplifies the treatment to consider the {\em excess} fragment spins,
defined as the amounts by which the spins deviate from 
what is required for the overall rigid rotation,
namely $\bld{s}_F \equiv \SF - \SF^{\rm rigid}$.
The rotational energies can then be written as the amounts associated
with the rigid rotation plus the contributions from the excess spins,
\beqar
E_\para &=& {J_\para^2\over2\Iz}
	+{(s_A^\para)^2\over2\IA^\para}+{(s_B^\para)^2\over2\IB^\para}\
=\ E_\para^{\rm rigid} + {s_{\rm twst}^2\over2\Itwst}\ ,\\ \nonumber
E_\perp &=& {J_\perp^2\over2\It}
	+{(s_A^\perp)^2\over2\IA^\perp}+{(s_B^\perp)^2\over22IB^\perp}
	+{(\bld{s}_A^\perp+\bld{s}_B^\perp)^2\over2\IR}\\ 
&=& E_\perp^{\rm rigid} 
    +{s_{\rm wrig}^2\over2\Iwrig} +{s_{\rm bend}^2\over2\Ibend}\ .
\eeqar
The last expressions give the excess energies in terms of the normal modes
(wriggling, bending, and twisting \cite{modes}), 
\beqar\label{wrig}
\bld{s}_{\rm wrig} &=& \bld{s}_A^\perp +\bld{s}_B^\perp,\,\
\Iwrig ={\IR\over\It} (\IA^\perp+\IB^\perp),\\ \label{bend}
\bld{s}_{\rm bend} &=& \Ibend \left({\bld{s}_A^\perp\over\IA^\perp}
	 -{\bld{s}_B^\perp\over\IB^\perp}\right),\,\
\Ibend ={\IA^\perp\IB^\perp \over \IA^\perp+\IB^\perp},\,\,\, \\ \label{twst}
s_{\rm twst} &=& \Itwst \left({\bld{s}_A^\para\over\IA^\para}
       -{\bld{s}_B^\para\over\IB^\para}\right),\,\
\Itwst ={\IA^\para\IB^\para\over\IA^\para+\IB^\para} .
\eeqar
The reduction factor $\IR/\It$ in the wriggling moment of inertia
is due to angular momentum conservation.
Because of the symmetry around \R, 
the perpendicular normal modes are doubly generate.
Thus there are a total of five distinct normal modes
which will be labeled by $m$.

In thermal equilibrium the ensemble average of the excess spins $s_m$ vanish,
$\langle s_m \rangle_T=0$, and so do their covariances, while
their variances are $\tilde{\sigma}_m^2=\langle s_m^2 \rangle_T={\cal I}_mT$.
(The tilde denotes the thermal equilibrium value.)
The equilibrium values for the excess fragment spins can be obtained
by use of the inverse relations.
The mean values vanish, $\langle\bld{s}_F\rangle=\bld{0}$.
Furthermore, the parallel components are generally perfectly anti-correlated, 
$s_B^\para=-s_A^\parallel$, so
\beq
\tilde{\sigma}_{AA}^\para =\tilde{\sigma}_{BB}^\para =\Itwst T,\,\ 
\tilde{\sigma}_{AB}^\para=-\Itwst\, T .
\eeq
Finally, the equilibrium fluctuations
of the fragment spin components in any perpendicular direction are
\beq
\tilde{\sigma}_{FF}^\perp=\left(1-{\IF^\perp\over\It}\right)\IF^\perp T,\,\
\tilde{\sigma}_{AB}^\perp=-{\IA^\perp\IB^\perp\over\It}T .
\eeq
Because $\It\gg\IF^\perp$ the relative fragment motion acts like an
angular-momentum reservoir, causing the reduction factor in the variance
to be close to unity and the anti-correlation between the two spins
to be relatively small.

\section{Spin transport treatment}
\label{FP}

In the preceding section, we have discussed the various angular-momentum bearing modes in the dinuclear system. 
In the course of the evolution from saddle to scission,
the two parts of the system exchange nucleons continually
and this mechanism changes their angular momenta in a stochastic manner.

Thus the above Langevin simulation of the shape parameters  $(R,c,\alpha)$
could be expanded to encompass the evolution of the excess fragment spins 
$(\sA,\sB)$.
However, it follows from the above discussion that, for a given shape and time,
the correlated spin-spin distribution 
has a standard mutivariate Gaussian form
(contrary to the distribution of the shape parameters).
Such a transport process can conveniently be treated 
within the Fokker-Planck framework \cite{NPA327,NPA383}
Accordingly, 
the temporal evolution of the distribution of the correlated excess spins,
$P(\sA,\sB)$, is governed by
\beq
{\del\over\del t}P = -\sum_i{\del\over\del s_i}V_i P\
+ \sum_{ij}{\del^2\over\del s_i\del s_j}D_{ij} P\ ,
\eeq
where $s_i$ represents any one of the six excess fragment spin components,
and it suffices to calculate just the mean evolutions $\overline{\s}_F(t)$ 
and the associated covariance matrix $\bld{\s}(\sA,\sB;t)$.

In accordance with the Einstein relation,
the transport coefficients (\ie\ the drift coefficients $V_i$
and the diffusion coefficients $D_{ij}$) are given in terms of 
the mobility tensor $\bld{M}(\sA,\sB)$ as follows,
\beq
V_i(\sA,\sB) = \sum_j M_{ij}f_j\ ,\,\,\
D_{ij}(\sA,\sB) = M_{ij} T\ ,
\eeq
where $f_i\equiv-\del E_{\rm rot}/\del s_i$ is the generalized force.
In the present case, $\bld f$ is linear in the excess spin components.

The ensemble average value of a spin component,
\beq
\overline{s}_i = \langle s_i\rangle = \int s_i\, P(\sA,\sB)\, d^3\sA d^3\sB ,
\eeq
is then governed by the following drift equation,
\beq
{\del\over\del t} \overline{s}_i 
= \int s_i {\del P\over\del t}\, d^3\sA d^3\sB 
= -\sum_j\nu_{ij}(\overline{\bld s})\, \overline{s}_j ,
\eeq
where the restoring coefficients are given by
\beq
\nu_{ij}\equiv -{\del V_i \over \del s_j}
= -\sum_kM_{ik} {\del f_k \over \del s_j}
= \sum_k M_{ik} {\del^2 E_{\rm rot} \over \del s_k\del s_j} .
\eeq

Furthermore, the covariance between the fluctuations 
of two spin components,
\beq
\sigma_{ij} = \langle S_i S_j\rangle - \overline{S}_i \overline{S}_j
	    = \langle s_i s_j\rangle - \overline{s}_i \overline{s}_j
	    = \langle s_i s_j\rangle\ ,
\eeq
evolves according to the following diffusion equation,
\beq\label{diffuse}
{\del\over\del t} \sigma_{ij} = 2D_{ij}(\overline{\bld s})
-\sum_k[\nu_{ik}(\overline{\bld s})\sigma_{kj}+\nu_{jk}(\overline{\bld s})\sigma_{ki}]\ .
\eeq
If the initial distribution is initially narrow, 
the covariance evolves as $\sigma_{ij}(t)\approx2D_{ij}t$ early on.
This linear growth is being counteracted by the increasing restoring term,
thereby ensuring that the equilibrium covariance is approached,
$\sigma_{ij}(t) \to\ \tilde{\sigma}_{ij}$.

\subsection{Nucleon exchange}

The present study considers the scenario where the dissipation
mechanism is multiple transfer of individual nucleons
between the two partners in a dinuclear system.
In the nucleon-exchange transport treatment \cite{NPA327,NPA383},
the mobility coefficients are proportional to the one-way
momentum current between the two binary partners,
\beq
{\cal N} = \quart m\rho\bar{v}\,\pi c^2\ ,
\eeq
the same quantity that governs the strength of the one-body window dissipation
used in the Langevin simulations of the shape evolution.
Here $m$ is the nucleon mass,
$\rho$ is the density of the nucleons in the bulk of the system,
and $\bar{v}\approx\mbox{$3\over4$}v_F$ is their mean speed.
The effective area over which the transfers can occur is
approximated by the window cross section, $\pi c^2$,
where $c(t)$ is the evolving neck radius
provided by the Langevin calculations.

When an individual nucleon is transfered,
all of the normal rotational modes are generally affected
but to different degrees depending on the specifics of the transfer.
These effects can be calculated by elementary means in a classical picture
\cite{NPA327}.

In order for a transfer to affect the twisting mode,
whose angular momentum is along the dinuclear axis,
the transfer must occur away from the axis, $\bld{r}=(x,y,0)$, and 
have a momentum component in the window plane that is perpendicular to $\bld r$.
It then causes equal and opposite changes
of the parallel excess fragment spins.
The associated mobility tensor for the parallel excess fragment spins
$s_A^\para$ and $s_B^\para$ is therefore given by
\beq
\bld{M}_{AB}^\parallel = {\cal N}\left(\begin{array}{rr}
c_0^2 & -c_0^2 \\ -c_0^2 & c_0^2 \end{array}\right)
= M_{\rm twst}\left(\begin{array}{rr} 1 & -1 \\ -1 & 1 \end{array}\right) ,
\eeq
where $c_0$ is the average transverse distance of the transfer from the axis,
$c_0^2=\half c^2$, and
$M_{\rm twst}={\cal N}c_0^2$ is the twisting mobility coefficient.

Somewhat similarly, if the nucleon transfers away from the axis, 
$\bld{r}=(x,y,0)$, with a momentum component $p_\para$ along the axis
it causes opposite changes
to the perpendicular components of the fragment excess spins.

More effective are transfers that have a momentum component, $p_\perp$,
perpendicular to $\hat{\bld R}$ because their angular momenta
(w.r.t.\ the fragment centers) are considerably larger,
namely $ap_\perp$ and $bp_\perp$, respectively,
where $a$ and $b$ are the distances from the two fragment centers
to the window (so $a^2, b^2 \gg c_0^2$).
Such transfers cause parallel changes in the perpendicular 
components of the fragment excess spins.
Thus the mobility tensor for the fragment spin components
in a perpendicular direction, $s_A^\perp$ and $s_B^\perp$, is given by
\beq
\bld{M}_{AB}^\perp = {\cal N}\left(\begin{array}{cc}
a^2+c_0^2 & ab-c_0^2 \\ ab-c_0^2 & b^2+c_0^2 \end{array}\right).
\eeq

It is instructive to reexpress this mobility tensor
in terms of wriggling (+) and bending (-),
\beq
\bld{M}_{\pm}^\perp = {\cal N}\left(\begin{array}{cc}
R^2 & R\delta \\ R\delta & \delta^2+c_0^2 \end{array}\right) .
\eeq
Here $\delta\equiv(a\IB-b\IA)/(\IA+\IB)$ is a measure of the asymmetry
and it enters because the transfer mechanism mixes the two normal modes
when the system is not symmetric.
While $\delta$ is smaller than the neck radius during most of the shape 
evolution, it dominates near scission where the neck radius shrinks to zero.

So it follows that wriggling and bending remain nearly independent
during the descent from saddle to scission.
Because generally $R^2\gg c_0^2$ the timescale for agitating wriggling
is considerably shorter than that for bending.

\section{Individual events}
\label{events}

After describing the Fokker-Planck transport treatment
of the excess fragment spins,
we now turn to the specific spin evolutions in individual fission events.
This will provide instructive insight that will facilitate 
the subsequent analysis of entire event ensembles.

The coupled first-order differential equations (\ref{diffuse})
for the spin covariances can be adequately solved by direct propagation, 
$\s_{FG}(t+\Delta t)\doteq\s_{FG}(t) + \dot{\s}_{FG}\Delta t$,
without the need to invoke a more advanced method.

\subsection{Parallel spin components}
\label{para}

It is simplest to consider the spin components parallel to the
dinuclear axis because they are one-dimensional and do not couple
to the orbital motion.

\begin{figure}[tbh]	    
\includegraphics[trim={2.5cm 0.2cm 8cm 8.0cm},clip,width=0.28\textwidth,
     angle=-90]{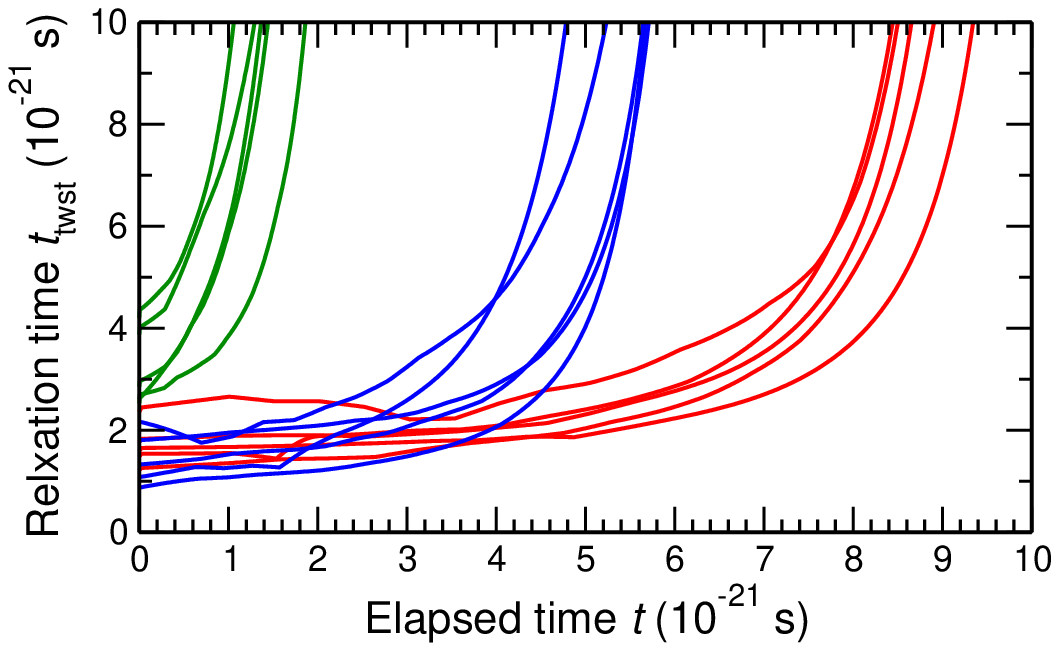}      
\caption{\label{f:tTwst}
The time dependence of the local twisting relaxation time
$t_{\rm twst}(t)=1/\nu_{\rm twst}(t)$ for the evolutions
shown in Fig.\ \ref{f:RcT}.}
\end{figure}

Because the driving force for twisting is given by
$f_{\rm twst}\equiv-\del(s_{\rm twst}^2/2{\cal I}_{\rm twst})/\del s_{\rm twst}
=-s_{\rm twst}/\Itwst$, 
the drift coefficient is $V_{\rm twst}=-M_{\rm twst}s_{\rm twst}/\Itwst$.
Thus the restoring rate becomes $\nu_{\rm twst}=M_{\rm twst}/\Itwst$.
The corresponding relaxation time, $t_{\rm twst}=1/\nu_{\rm twst}$,
is shown in Fig.\ \ref{f:tTwst} for the evolutions shown in Fig.\ \ref{f:RcT}.
As long as the shape meanders around in the diffusive regime near the saddle,
where the system is fairly compact and the neck is wide open,
$t_{\rm twst}$ remains rather short.
But as scission is approached and the neck closes,
$t_{\rm twst}$ rapidly diverges, 
effectively halting the evolution of the mode.

If the geometry were to remain constant,
the mean value of the twisting spin would decay exponentially,
$\overline{s}_{\rm twst}(t)=\overline{s}_{\rm twst}(0)\exp(-t/t_{\rm twst})$,
with the decay time being $t_{\rm twst}$,
while the variance would evolve as
\beq
\s_{\rm twst}^2(t) = \s_{\rm twst}^2(0)\ e^{-2t/t_{\rm twst}}
	 +[1-e^{-2t/t_{\rm twst}}]\, \tilde{\s}_{\rm twst}^2 .
\eeq
Thus, with a time scale of $\half t_{\rm twst}$,
the initial fluctuations would subside (first term)
as thermal equilibrium is approached (second term).

\begin{figure}[tbh]	    	
\includegraphics[trim={2.5cm 0.3cm 5.5cm 0cm},clip,width=0.25\textwidth,
     angle=-90]{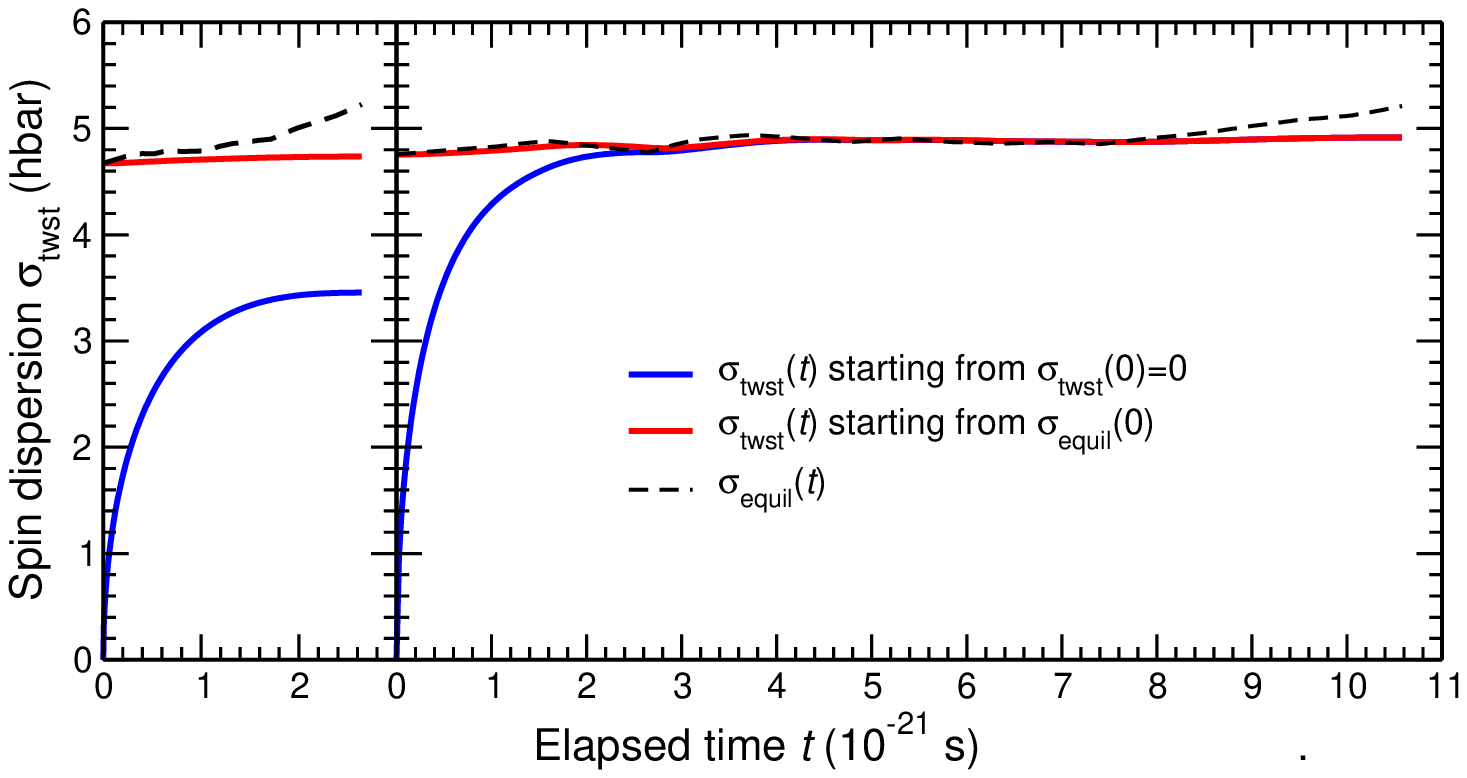}      
\caption{\label{f:sTwst}
The time evolution of the twisting mode for 
a fast (left panel) and a slow (right panel) event.
The mode is initially either in thermal equilibrium (red), 
$\sigma_{\rm twst}(0)=\tilde{\sigma}_{\rm twst}(0)$,
or not populated at all (blue), $\sigma_{\rm twst}(0)=0$, and it is then
followed until scission by solving the diffusion equation (\ref{Dtwst}).
Also shown is the evolving local equilibrium dispersion (dashed),
$\tilde{\sigma}_{\rm twst}(t)=[\Itwst(t)T(t)]^{1/2}$.}
\end{figure}

In the present study, the geometry is given by the Langevin simulations, 
as described above.
The evolution equation for the twisting variance must then be solved numerically
(which is straightforward).
The resulting development of $\s_{\rm twst}^2$ is shown in Fig.\ \ref{f:sTwst}
for a fast and a slow event.
The mode is initially either fully equilibrated,
$\sigma_{\rm twst}(0)=\tilde{\sigma}_{\rm twst}(0)={\cal I}_{\rm twst}(0)T(0)$,
or entirely unpopulated, $\sigma_{\rm twst}(0)=0$.

For the slow event, the saddle-to-scission time is considerably larger
than the relaxation time.
Therefore, when initially in equilibrium the mode remains
close to the local equilibrium, and when initially unpopulated 
it has time to equilibrate well before scission is approached,
$\s_{\rm twst}(t)\approx\tilde{\s}_{\rm twst}(t)$.
But, as the neck radius shrinks relatively rapidly near scission,
the relaxation time grows large and the mode is effectively frozen in,
while the local equilibrium fluctuations, 
as given by the dispersion $\tilde{\s}_{\rm twst}(t)$,
grow rapidly due primarily to the growth of the temperature $T(t)$
(see Fig.\ \ref{f:RcT}).

The fast event reaches the scission region rapidly,
so the divergence of the relaxation time starts quickly.
 As a result, when initially equilibrated the mode quickly 
falls behind the growing local equilibrium fluctuations
and when initally unpopulated the mode does not have time to equilibrate
before the evolution is frozen in due to the shrinking neck.

Thus the final fluctuations of the twisting amplitude
will always be smaller than the equilibrium fluctuations at scission,
and the more so the shorter the saddle-to-scission time is.

The parallel components of the excess fragment spins,
are given in terms of the twisting amplitude as 
$s_A^\para=s_{\rm twst}$ and $s_B^\para=-s_{\rm twst}$.
Hence their average values are governed by
$d\overline{s}_F^\para(t)/dt = - \overline{s}_F^\para(t)/t_{\rm twst}(t)$,
while the evolution equation for the second moments 
$\s_{FG}^\para=\langle s_F^\para s_G^\para\rangle$ is
\beq\label{Dtwst}
{d\over dt} \s_{FG}^\para(t) 
	= 2M_{FG}^\para(t) T(t) - 2\s_{FG}^\para(t)/t_{\rm twst}^\para(t)\ ,
\eeq
yielding $\s_{AA}^\para(t) =\s_{BB}^\para(t)
=-\s_{AB}^\para(t) =\s_{\rm twst}^2(t)$.

\subsection{Perpendicular spin components}
\label{perp}

The perpendicular spin components present a richer picture
because there are two different normal modes, wriggling and bending.
In addition, each one is doubly degenerate because of the axial symmetry.
But all the perpendicular directions behave similarly,
so it suffices to consider just one (arbitrary) perpendicular direction.

The generalized perpendicular driving force is
\beq
f_F^\perp \equiv -{\del E_\perp \over\del s_F^\perp}
= -{s_F^\perp\over\IF^\perp} - {s_A^\perp+s_B^\perp \over\IR} ,
\eeq
so the restoring rate tensor becomes
\beq
\bld{\nu}^\perp = {\cal N}\left(\begin{array}{cc}
{a^2+c_0^2\over\IA} + {aR\over\IR} & {ab -c_0^2\over\IB} + {aR\over\IR} \\ 
{ab -c_0^2\over\IA} + {bR\over\IR} & {b^2+c_0^2\over\IB} + {bR\over\IR}
\end{array}\right) .
\eeq
The drift equation governing the mean evolution 
of the perpendicular excess fragment spins is then 
\beq\label{VAB}
{d\over dt}{\overline{s}}_F^\perp
= -\sum_G \nu_{FG}^\perp(\overline{s}_F)\, \overline{s}_G^\perp\ .
\eeq
However, the drift equation is of little interest 
because the scenarios considered usually start from excess spin distributions 
that have zero mean values, 
$\bar{\bld s}_F(0)=\bld{0}$, such as thermal equilibrium,
so they will remain zero.

The focus of the present study is on the dynamical evolution of the associated
covariance tensor $\bld{\s}^\perp$ having the elements $\s_{FG}^\perp\equiv
\langle S_F^\perp S_G^\perp\rangle-\overline{S}_F^\perp \overline{S}_G^\perp =
\langle s_F^\perp s_G^\perp\rangle-\overline{s}_F^\perp \overline{s}_G^\perp$.
The diffusion equation governing the evolution of $\bld{\s}^\perp$ is
\beq\label{DAB}
{d\over dt}{\s}_{FG} ^\perp= 2M_{FG}^\perp T
-\sum_H\left[\nu_{FH}^\perp\s_{HG}^\perp
	    +\nu_{GH}^\perp\s_{HF}^\perp\right] .
\eeq

\begin{figure}[b]	    
\includegraphics[trim={2.5cm 0.3cm 5.5cm 0cm},clip,width=0.25\textwidth,
     angle=-90]{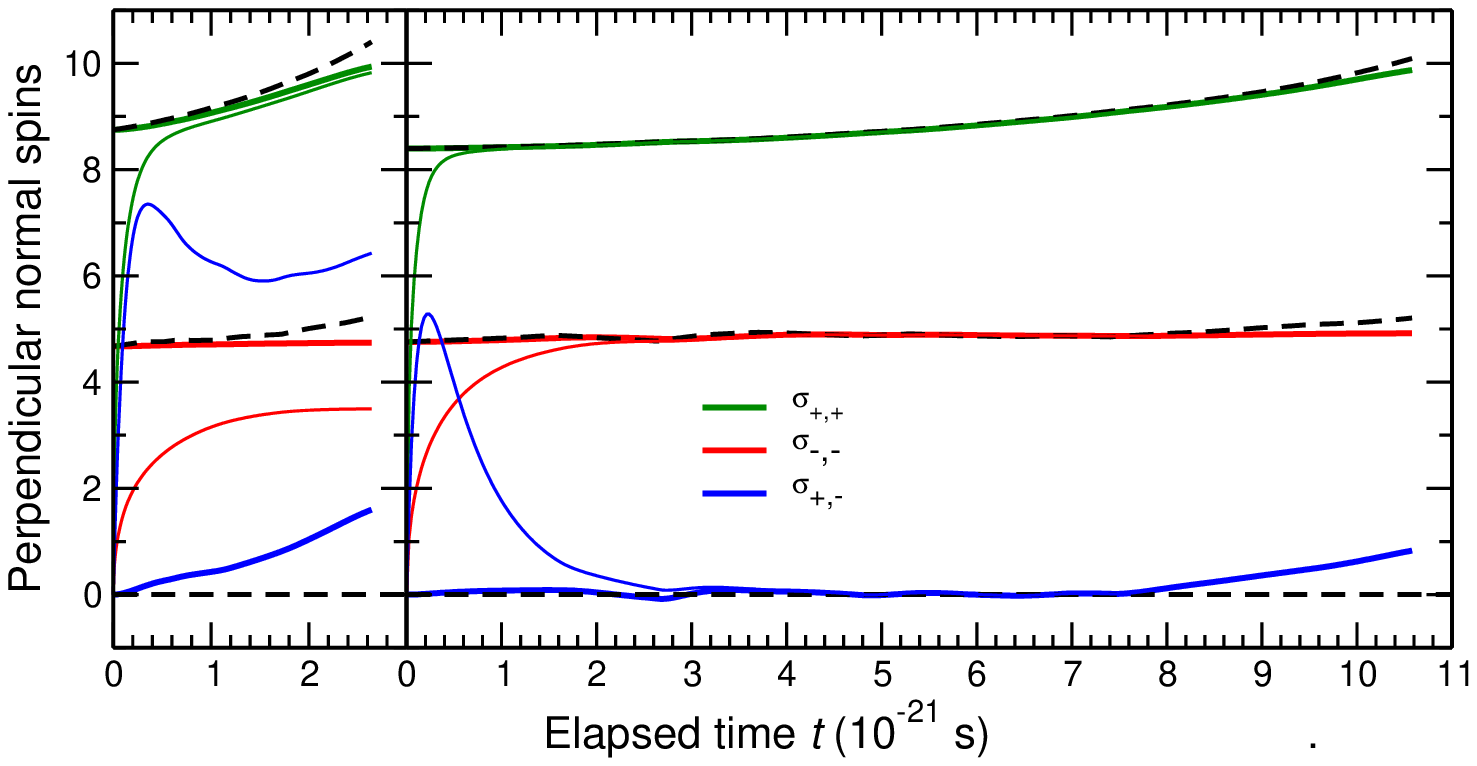}      
\caption{\label{f:spm}
The time evolution of the normal spin modes
for the two events considered in Fig.\ \ref{f:sTwst},
the dispersion of the wriggling (green) and the bending (red)
amplitudes, $\sigma_+$ and $\sigma_-$, 
as well as their covariance $\sigma_{+,-}$ (blue).
The evolving local equilibrium dispersions are also shown (dashed black).
The initial normal spin amplitudes are either in 
statistical equilibrium (thick curves) or vanish (thin curves).
}\end{figure}		   

It is instructive to first view the spin evolution
in terms of the two perpendicular normal modes, wriggling and bending.
Being normal modes, wriggling and bending are uncorrelated in equilibrium
but when the system is asymmetric the nucleon transfers mix the two modes,
leading to a fast and a slow dynamcial eigenmode with
relaxation times given approximately by
$t_{\rm fast}^{-1}\approx{\cal N}R^2/{\cal I}_+$ and
$t_{\rm slow}^{-1}\approx{\cal N}c_0^2/{\cal I}_-=t_{\rm twst}^{-1}$.
Because the mixing is relatively weak,
these modes are approximately identical to the wriggling and bending modes.

The evolution of the dynamically coupled normal modes
is illustrated in Fig.\ \ref{f:spm}
for the same Langevin events as considered in Fig.\ \ref{f:sTwst}.
It can be seen how the wriggling mode relaxes very quickly 
and achieves approximate euilibrium also in the fast event,
even if initially unpopulated.
By contrast, when initially unpopulated, 
the bending mode achieves only partial equilibration in the fast event.

The covariance between the two normal modes remains small
when the initial distributions have the equilibrium form.
But when the modes are initially unpopulated, 
the fact that the time scale for wriggling is much shorter
than that for bending causes their correlation 
to first be dominated by wriggling-inducing transfers,
which generate parallel fragment spins, 
and the relaxation towards uncorrelated normal spins occurs only later
due to the slower action of the bending-inducing tranfers 
that generate opposite fragment spins.
For the fast event, only the inital growth of the correlation occurs
before scission halts the evolution.

When scission is approached and the neck radius shrinks,
the fluctuations of the two normal modes cannot keep up with the
increasing equilibrium fluctuations, as was the case for the twisting mode,
so they freeze out at somewhat smaller values.
It should also be noted that 
the correlation between the two normal modes grows positive because
bending-inducing transfers (being proportional to $c^2(c_0^2+\delta^2)$) are
impeded faster than wriggling-inducing transfers 
(which are proportional to $c^2R^2$).

\begin{figure}[tbh]	    
\includegraphics[trim={2.5cm 0.3cm 5.5cm 0cm},clip,width=0.25\textwidth,
     angle=-90]{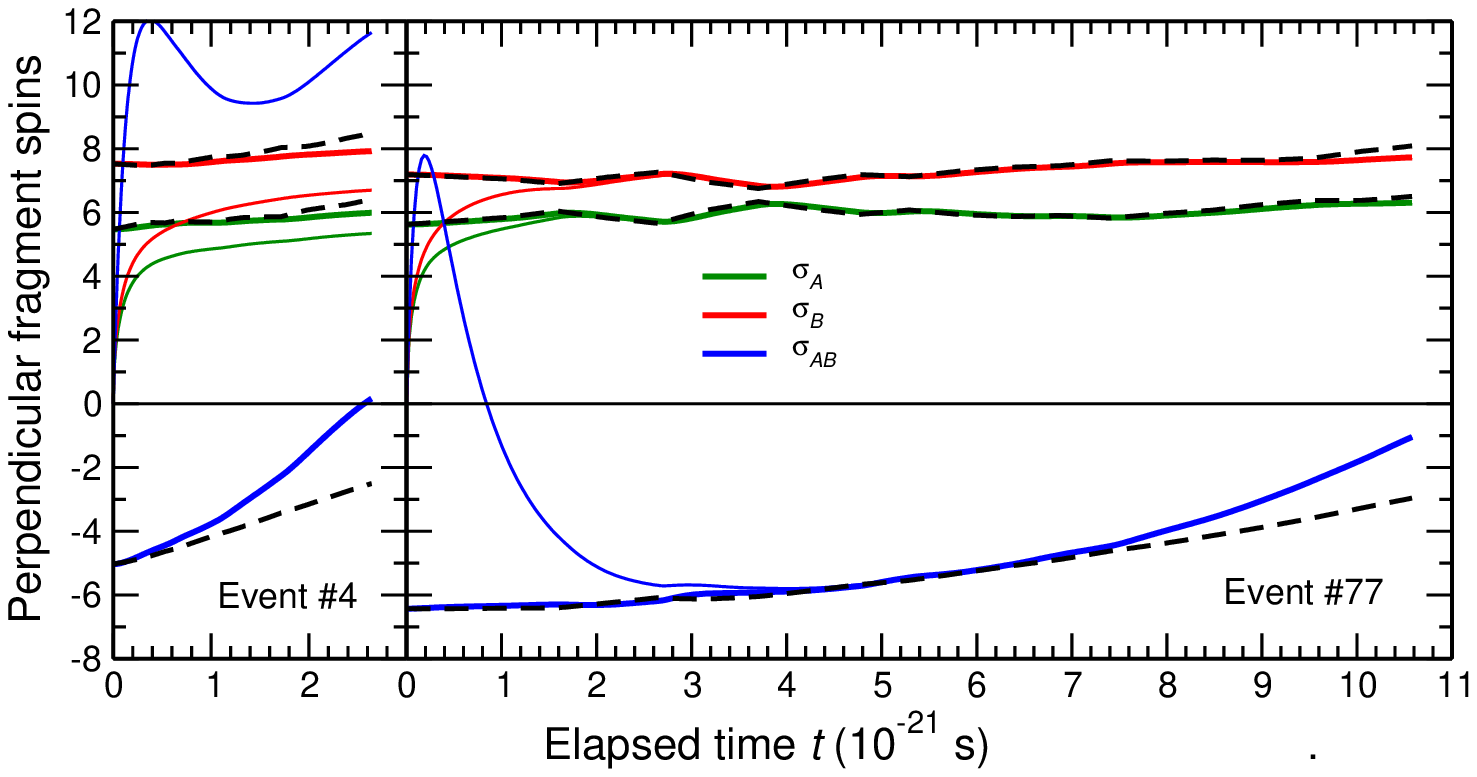}      
\caption{\label{f:sAB}
The time evolution of the fragment spin components in an (arbitrary)
direction perpendicular to \R\
for the fast and slow events shown in Figs.\ \ref{f:sTwst} and \ref{f:spm}.
The solid curves show the evolving fragment spin dispersions,
$\sigma_A^\perp$ (green) and $\sigma_B^\perp$ (red), 
as well as their covariance $\sigma_{AB}^\perp$ (blue),
together with the local equilibrium values (dashed black).
The initial correlated excess spin fluctuations are either in 
statistical equilibrium (thick curves) or vanish (thin curves).
}\end{figure}		   

The features discussed above are reflected in the behavior of the individual 
perpendicular fragment spins,
as shown in Fig.\ \ref{f:sAB} for the same two events.

For the slow event, the covariance tensor for the initially equilibrated case
remains close to the local equilibrium tensor $\tilde{\bld{\s}}^\perp(t)$,
until scission is approached, and the covariance tensor for the initially
fluctuation-free case has time to equilibrate well before then.
But as scission is approached and the mobility coefficients drop off,
the fluctuations are effectively frozen in and cannot adjust to the
still evolving local equilibrium fluctuations.

The correlation coefficient $c_{ij}\equiv\s_{ij}/[\s_i\s_j]$ provides 
a dimensionless measure of the degree of correlation between two quantities.
As noted in Sect.\ \ref{spins}, the two perpendicular fragment spin components
are only slightly (anti)correlated because $\IR\gg\IF$,
so the thermal equilibrium value
of their correlation coefficient is rather small,
$c_{AB}^\perp\approx -[\IA\IB]^{1/2}/\IR \approx -5\%$.
The late growth spurt of $\sigma_{AB}^\perp(t)$,
due to the dominance of wriggling agitations near scission,
brings its final value closer to zero (see Fig.\ \ref{f:sAB}).
Consequently, the correlation coefficient, already small in equilibrium,
ends up being even smaller at  scission,
indicating that the perpendicular excess spin components
of the emerging fragments are practically uncorrelated,
as was also found experimentally \cite{Nature590}.

For the fast event, the saddle-to-scission time is not much larger than 
the relaxation time for the bending mode
which therefore does not reach equilibrium if initially unpopulated
(see Fig.\ \ref{f:spm}).
As a consequence, the individual fragment spins 
end up with magnitudes significantly below their equilibrium values.
When starting from equilibrium, the covariance between the
two perpendicular spin components exhibits a rapid growth 
similar to the late behavior in the slow event.
In the (less realistic) scenario when there are no initial fluctuations, the
early behavior is similar to what occurs for that scenario in the slow event:
a rapid growth of $\s_{AB}^\perp(t)$ to large positive values.
But the subsequent relaxation towards the local equilibrium covariance,
caused by the (slower) action of bending agitations,
is ineffective because the shape is already near scission
so those are being impeded and $\s_{AB}^\perp(t)$ then exhibits a 
late growth spurt similar to what occurs in the slow event.

\section{Ensemble results}
\label{ensembles}

After the typical behavior of the spin distribution 
in individual shape evolutions was illustrated above,
entire ensembles of spin distributions are now analyzed.
As desrcribed in Sect.\ \ref{shapes},
three fissioning systems, one of them at two different excitations,
are considered, using two different strengths of the one-body wall friction.

It should be noted that all of the fission cases considered in this study
are at fairly high excitation energy.  
In this energy regime the microscopic effects are 
small and they may be neglected in the simulation of the shape evolution,
making it practically possible to generate large samples of fission events.
Furthermore, in order to avoid complicating the picture,
the calculation does not include the possibility of neutron evaporation 
during the shape evolution.
As a result, the evolution proceeeds at a relatively high temperature
and the calculated fragment spins are correspondingly large.
However, the effects seen are expected to be general
and thus they should also be present at lower energies.

The spin observables extracted from the considered eight ensembles
are discussed below and summarized in Table \ref{t:I}.

\begin{table}[bth]		 
\begin{tabular}{ccc|rcccc}\hline\hline\\[-2ex]
$^{Z_0\!\!}A_0$ & $E^*$ & $k_s$ &
~~~\raisebox{1.0ex}{$\langle t_{\rm ss}\rangle$} 
\hspace{-2.5em} \raisebox{-1.2ex}{$\pm\s_{\rm ss}$}~
 &
~~$\s_A$~~~ & ~~$\s_B$~~~ & ~~$c_{AB}$~~~ & $~~\theta_F~~$ \\[1ex]
\hline\\[-2.5ex]
$^{182}$Hg & 46 & 1     &      4.06 & 5.04 & 6.16 & -3.1 & 63.4\\[-0.6ex]
~ & ~ & ~               & $\pm$2.20 & 5.32 & 6.56 & -6.0 & 63.1\\[0.5ex]
$^{182}$Hg & 46 & 0.25  &      2.41 & 4.92 & 6.10 & -4.8 & 63.3\\[-0.6ex]
~ & ~ & ~               & $\pm$1.03 & 5.12 & 6.39 & -6.5 & 63.1\\[0.5ex]
$^{202}$Po & 46 & 1     &      5.30 & 5.51 & 6.67 & -2.6 & 63.5\\[-0.6ex]
~ & ~ & ~               & $\pm$2.63 & 5.82 & 7.10 & -5.9 & 63.0\\[0.5ex]
$^{202}$Po & 46 & 0.25  &      2.98 & 5.36 & 6.59 & -4.4 & 63.3\\[-0.6ex]
~ & ~ & ~               & $\pm$1.23 & 5.60 & 6.93 & -6.5 & 63.0\\[0.5ex]
$^{236}$U & 46 & 1      &      6.89 & 6.34 & 7.51 & -2.9 & 63.5\\[-0.6ex]
~ & ~ & ~               & $\pm$3.06 & 6.70 & 8.11 & -5.7 & 63.0\\[0.5ex]
$^{236}$U & 46 & 0.25   &      3.58 & 6.12 & 7.53 & -3.5 & 63.5\\[-0.6ex]
~ & ~ & ~               & $\pm$1.30 & 6.42 & 7.96 & -6.3 & 63.0\\[0.5ex]
$^{236}$U & 70 & 1      &      5.94 & 6.90 & 8.52 & -3.7 & 63.4\\[-0.6ex]
~ & ~ & ~               & $\pm$2.82 & 6.42 & 7.96 & -5.8 & 63.1\\[0.5ex]
$^{236}$U & 70 & 0.25   &      3.21 & 6.72 & 8.46 & -5.0 & 63.4\\[-0.6ex]
~ & ~ & ~               & $\pm$1.24 & 6.94 & 8.59 & -6.3 & 63.1\\[0.5ex]
\hline\hline
\end{tabular}
\caption{For the various cases considered 
(the fissioning system $^{Z_0}A_0$, its excitation energy $E^*$ (MeV),
and the dissipation parameter $k_s$) are shown
the mean saddle-to-scission time $\langle t_{\rm ss}\rangle$
and its dispersion $\s_{\rm ss}$ (in zs), as well as
the dispersions of the fragment spin distributions,
$\s_A$ and $\s_B$ ($\hbar$),
together with the associated correlation coefficient $c_{AB}$ (\%)
and the average orientation angle $\theta_F$ ($^\circ$).
For each case, the first line shows the results of the dynamical calculation,
while the second line shows the corresponding equilibrium values at scission.
}\label{t:I}
\end{table}		     	

\subsection{Fragment spin magnitudes}

The excess fragment spins ${\bld s}_F$ may have components
both parallel and perpendicular to the fission axis,
$s_F^\para$ and ${\bld s}_F^\perp$.
The variance of the parallel component
is given by the twisting variance, as discussed above in Sect. \ref{para}.
But because the discussion above in Sect.\ \ref{perp}
was concerned with the spin components in just one of the two perpendicular
directions, the covariance elements for the total perpendicular spin components
are twice as large as those considered there.
Thus the total variance of the excess fragment spin ${\bld s}_F$ is
\beq \s_F^2\ =\ (\s_F^\para)^2\ +\ 2(\s_F^\perp)^2 \eeq
and the dispersion $\s_F$ is a measure of the mean spin magnitude
in the given evolution.

\begin{figure}[tbh]	    
\includegraphics[trim={3.2cm 0.3cm 7.7cm 0cm},clip,width=0.26\textwidth,
     angle=-90]{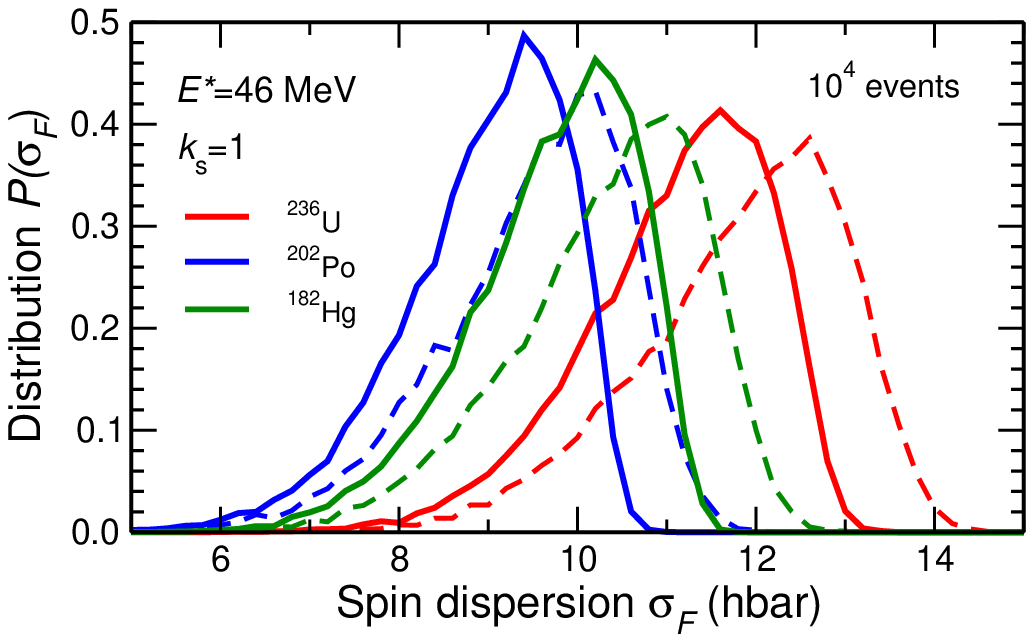}      
\caption{\label{f:SF}
The distribution of the fragment spin dispersion $\s_F$ at scission
for the three cases with $E^*=46\,\MeV$, calculated with $k_s=1$.
For each case, the Fokker-Planck results (solid) 
and the equilibrium results (dashed) are shown.}
\end{figure}

Figure \ref{f:SF} shows the distribution of the fragment spin dispersion,
$P(\s_F;t_{\rm ss})$, for various event ensembles,
as obtained by solving the Fokker-Planck transport equation for 
each individual Langevin shape evolution from the saddle region to scission.
For each event ensemble is also shown the distribution of the corresponding 
equilibrium dispersions at scission, $P(\tilde{\s}_F;t_{\rm ss})$.

As expected from the above discussion in Sect.\ \ref{events},
the dynamical fluctuations cannot adjust to 
the rapid growth of the equilibrium fluctuations near scission.
As a result, $P(\s_F)$ is shifted downwards by about one $\hbar$
relative to the equilibrium distribution.
This is a universal feature seen in all the considered cases.

The distributions shown in Fig.\ \ref{f:SF} were calculated for shape evolutions
generated with $k_s=1$ giving the standard strength of the wall dissipation
\cite{onebody}.
Those calculated for shape evolutions generated with the commonly advocated 
value $k_s=0.25$ are very similar, so the spin magnitude distribution
is not very sensitive to $k_s$.

Furthermore, as one would expect, an increase of
the nuclear excitation energy $E^*$ generally increases the spin fluctuations.
In particular, the spin magnitude distributions for fission of uranium 
at $E^*=70\,\MeV$ are shifted upwards by nearly 2$\hbar$
relative to those for $E^*=46\,\MeV$.

\begin{figure}[tbh]	    
\includegraphics[trim={3.2cm 0.3cm 7.7cm 0cm},clip,width=0.26\textwidth,
     angle=-90]{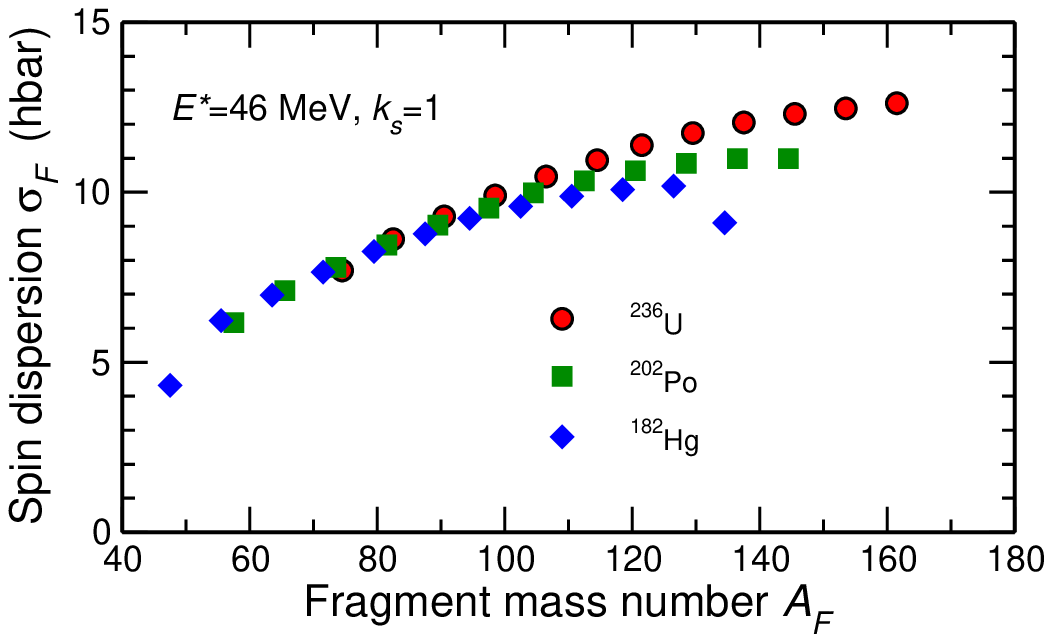}      
\caption{\label{f:SFm}
The mean spin dispersion for fragments in various mass ranges
for the three cases having $E^*=46\,\MeV$, calculated with $k_s=1$.
As suggested by Fig.\ \ref{f:SF},
the corresponding equilibrium values are about 10\%\ larger.}
\end{figure}

Because it is possible to experimentally gate on the fragment mass,
it is interesting to determine the mass dependence of the spin magnitude.
This is illustrated in Fig.\ \ref{f:SFm} for the three caases
that have $E^*=46\,\MeV$ and use $k_s=1$.
The corresponding calculations using $k_s=0.25$ yield very similar results.
The average spin dispersions for the light and heavy fragments
are listed in Table \ref{t:I} for the eight fission cases considered.

As one would expect, the heavier fragments tend to have larger spins,
but this trend gradually weakens, and more so for the lighter systems;
for $^{182}$Hg the mean spin magnitude 
even decreases for the heaviest fragments.
On  the other hand, the mean spins of the lighter fragments
are not noticeably dependent on the compound system.
It would thus seem that gating on the fragment mass
when measuring the fragment spins may be particularly informative.

\subsection{Fragment spin correlations}

The fragment spin fluctuations are generated by microscopic
processes (in this case the transfer of individual nucleons)
that change the spins in a correlated manner.
Therefore the resulting values of ${\bld s}_A$ and ${\bld s}_B$ 
are generally correlated,
as signalled by the non-zero value of the spin-spin covariance,
\beq
\s_{AB} \equiv \langle{\bld s}_A\cdot{\bld s}_B\rangle
= \s_{AB}^\para + 2\s_{AB}^\perp\ .
\eeq
It is convenient to consider the corresponding correlation coefficient,
$c_{AB}\equiv \s_{AB}/[\s_A\s_B]$;
its possible values 
range from +1 (fully correlated exces spins: ${\bld s}_A={\bld s}_B$)
to -1 (fully anti-correlated excess spins: ${\bld s}_A=-{\bld s}_B$).

It was noted above that the two fragment spins 
are slightly anti-correlated in thermal equilibrium,
an effect of angular-momentum conservation
that becomes ever smaller as the system elongates.
The spins are even less correlated when calculated dynamically,
due to the dominance of wriggling-inducing transfers near scission.

\begin{figure}[tbh]	    
\includegraphics[trim={3.2cm 0.3cm 7.7cm 0cm},clip,width=0.26\textwidth,
     angle=-90]{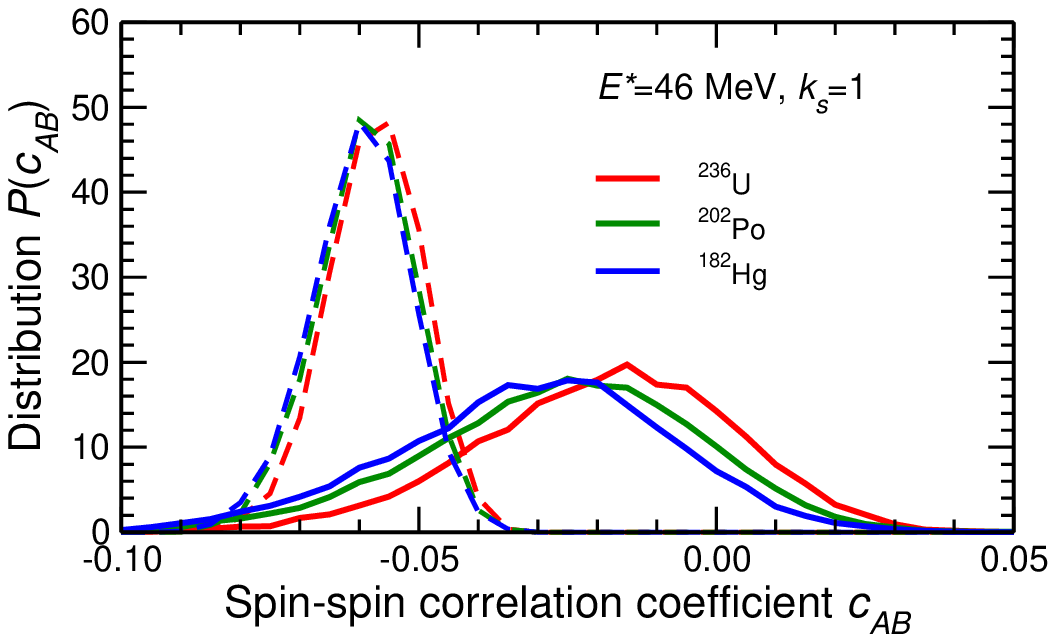}      
\caption{\label{f:cAB}
The distribution of the fragment spin correlation coefficient
at scission for fission of $^{236}$U, $^{202}$Po and $^{182}$Hg
at $E^*$=46~MeV with $k_s=1$,
as calculated in thermal equilibrium (dashed)
and as given by the Fokker-Planck evolution (solid).
}\end{figure}

These features are apparent in the distributions of the correlation coefficient,
as illustrated in Fig.\ \ref{f:cAB}.
The equilibrium value of the correlation coefficient is about -6\%
for all the cases considered.
When the spin distribution is followed dynamically 
within the Fokker-Planck framework the correlation distribution
is broader and its centroid is reduced by a factor of two to three.

When a weaker dissipation is used, $k_s=0.25$, the equilinrium distributions
of $c_{AB}$ are nearly the same, while the dynamical distributions 
are shifted by only half as much, approximately.
Unfortunately, all the cases yield correlations that are so small 
that it will likely be practically impossible to distinguish experimentally
between the statistical and the dynamical scenarios.

If $\psi_{AB}$ denotes the opening angle between ${\bld s}_A$ and ${\bld s}_B$
then ${\bld s}_A\cdot{\bld s}_B = s_As_B\cos\psi_{AB}$.
The average opening angle, $\overline{\psi}_{AB}$, 
is then approximately given by
\beq
\cos\overline{\psi}_{AB} \approx \langle\cos\psi_{AB}\rangle
\approx {\langle{\bld s}_A\cdot{\bld s}_B\rangle \over
 \left[\langle s_A^2\rangle \langle s_B^2\rangle\right]^{1/2}} = c_{AB}
\eeq
for each shape evolution.
So the correlation coefficient $c_{AB}$ gives a rough indication
of the average opening angle between the two fragment spins, 
$\overline{\psi}_{AB}\approx \arccos(c_{AB})$.
The ensemble average of $c_{AB}$ is given in Table \ref{t:I} 
for the various cases considered.

\subsection{Fragment spin orientation}

\begin{figure}[b]	    
\includegraphics[trim={3.2cm 0.3cm 7.7cm 0cm},clip,width=0.26\textwidth,
     angle=-90]{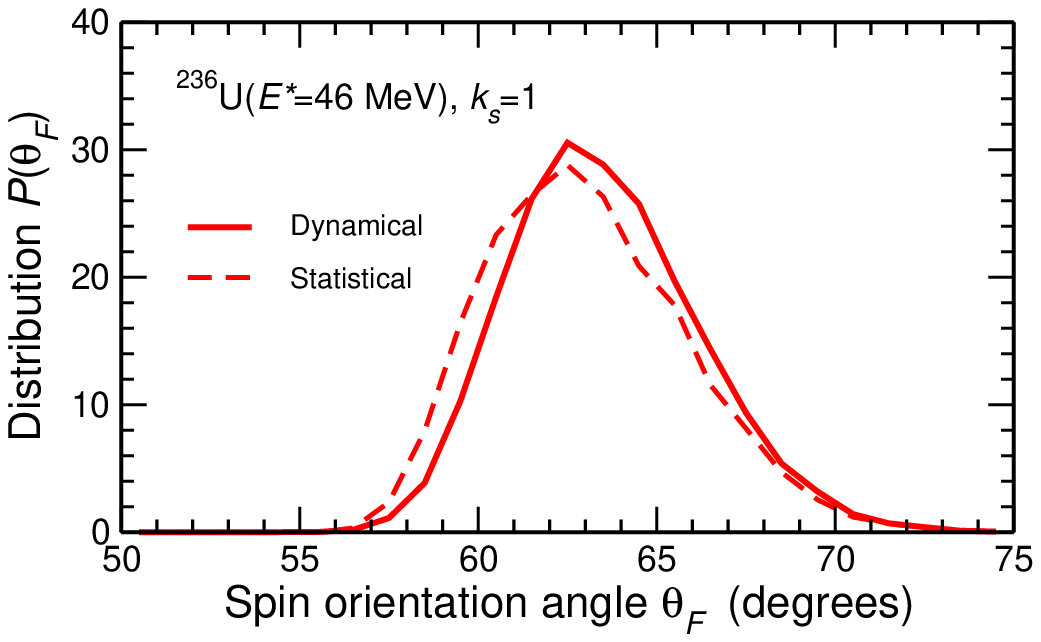}      
\caption{\label{f:thF}
The distribution of the orientation angle of the fragment spin 
relative to the fission direction, $\theta_F$,
as obtained by solving the Fokker-Planck equation
for the individual Langevin shape evolutions (solid)
and the corresponding equilibrium distribution (dashed).
}\end{figure}	            

The orientation of the fragment spin $\SF$ relative to the
dinuclear axis at scission, $\hat{\bld R}$, is given by the angle $\theta_F$
where $\cos\theta_F = \hat{\SF}\cdot\hat{\bld R}$.
In the present study, the contribution from rigid rotation 
is small in comparison with the typical excess spin,
so $\tan\theta_F = S_F^{\para}/S_F^{\perp} \approx s_F^{\para}/s_F^\perp$.
The ensemble average of $\tan^2\theta_F$ is therefore approximately given by
\beq
\langle\tan^2\theta_F\rangle \approx
{\langle (s_F^\perp)^2 \rangle \over \langle (s_F^\para)^2 \rangle}
= {2(\sigma_F^\perp)^2 \over (\sigma_F^\para)^2}\ ,
\eeq
where $\sigma_F^\perp$ is the fragment spin dispersion in one
of the perpendicular directions (see Sect.\ \ref{perp}).

The corresponding equilibrium value is
\beq
\langle\tan^2\theta_F\rangle_T
\approx {2(\tilde{\s}_F^\perp)^2 \over (\tilde{\s}_F^\para)^2}
= {2[1-(\IF^\perp/\Itot^\perp)]\IF^\perp \over 
  \IA^\para\IB^\para / (\IA^\para+\IB^\para)}\ ,
\eeq
where $\Itot^\perp=\IA^\perp+\IB^\perp+\IR$.
For equal touching spheres we have $\Itot=7\IF$,
so $\langle\tan^2\theta\rangle_T \approx \mbox{$24\over7$}$, and when 
$\Itot\gg\IF$ we have $\langle\tan^2\theta\rangle_T \approx 4$.
If we define the mean orientation angle $\bar{\theta}_F$ by
$\tan^2\bar{\theta}_F=\langle\tan^2\theta_F\rangle$, 
the equilibrium values of $\bar{\theta}_F$ are
$61.6^\circ$ and $63.5^\circ$, respectively.

The distribution of the spin orientation angle $\overline{\theta}_F$ 
is shown in Fig.\ \ref{f:thF}
for fission of $^{236}$U at $E^*=46\,\MeV$ with $k_s=1$.
The results for all the other cases considered are very similar.
It is notable that the dynamical distribution,
obtained by solving the Fokker-Planck equation for each shape evolution
in the ensemble, is very similar to the statistical distribution
obtained by using the thermal equilibrium variances at scission.
This result is thus very robust,
being nearly independent of the specific scenario considered.
It would therefore be particularly interesting to remeasure this quantity
with improved accuracy relative to the pioneering experiments
\cite{WilhelmyPRC5,WolfPRC13} 
that found the spin orientation angle to be consistent with $90^\circ$.

However, before confrontation with experimental data is made,
it should be borne in mind that the present calculations
are likely to underestimate the orientation angle.
This expectation is based on the fact that the most likely orientation angle 
is roughly determined by the ratio of the moments of inertia,
$\tan^2\overline{\theta}_F \approx 4\IF^\perp/\IF^\para$.
These have here been taken to be equal, whereas
 $\IF^\perp$ tends to be larger than $\IF^\para$ for typical fission fragments.
Thus, for example, if $\IF^\perp=2\IF^\para$ 
then the most likely orientation angle
is increased to $\overline{\theta}_F \approx 70^\circ$.

\section{Concluding remarks}

The present study has explored the consequences of the nucleon-exchange mechanism
for the dynamical build-up of fragment angular momenta in fission.
A well eststablished Langevin simulation model was used
to generate large samples of shape evolutions.
For the stage from saddle to scission,
we applied the nucleon-exchange transport model
developed previously for the description of damped nuclear reactions.
In this manner, without the neeed for introducing any parameters,
it was possible to calculate the saddle-to-scission development
of the six-dimensional spin distribution of the fledging fission fragments
for each separate shape evolution.

A number of general features were identified:
\begin{itemize}
\item{\em Magnitude.}
Because nucleon transfer is progressively impeded
as scission is approached while, at the same time,
the temperature is rising rapidly, 
the evolving spin distribution freezes out before scission,
leaving the fragments with spins that are somewhat smaller than would be
expected from statistical considerations at scission.

\item{\em Correlation.} 
Because the transfers agitating the negative normal modes
(bending and twisting) cease earlier than those agitating the positive
normal modes (wriggling), the final correlation between the spins
becomes even smaller than in equilibrium.

\item{\em Orientation.} 
The orientation of the fragment spin, 
whether calculated dynamically or statistically, 
is distributed around a value significantly below $90^\circ$.

\item{\em Mass dependence.} 
As brought out in Fig.\ \ref{f:SFm}, the calculations suggest that
the fragment-mass dependence of the average spin magnitudes 
evolves in a non-trivial manner with the size of the fissioning system
which is something that is amenable to experimental investigation.
\end{itemize}

It would obviously be interesting to have comparable studies
of alternative mechanisms
(such as those studied in Refs.\ \cite{ScampsPRC110}), so key differences
can be identified and suitable experimental tests undertaken.

It is instructive to compare the present results
with the expectations presented in Ref.\ \cite{RDV}.
Within the same physical picture, 
but without any shape evolution simulations or spin transport equations,
estimates were made of the mode relaxation times 
and the saddle-to-scission duration $t_{\rm ss}$.
The estimate $t_{\rm ss}\approx1-4\,\zs$ 
is a good representation of the durations obtained here with $k_s=0.25$;
those obtained with $k_s=1$ are nearly twice as long.
However, the results calculated here are not very sensitive to $k_s$,
nor to $t_{\rm ss}$, because the modes (including twisting)
tend to equilibrate well before scission.
The most noticeable non-equilibrium effect occurs as scission is approached
when the mobility coefficients decrease rapidly, and faster for twisting
and bending than for wriggling, 
a general effect that was not discussed in Ref.\ \cite{RDV}
and which, to our knowledge, has not been recognized earlier.
 
Furthermore, the expectation in Ref.\ \cite{RDV} that bending
grows more prominent at larger mass asymmetry is validated by the present 
dynamical calculations but the effect is only at the level of one percent.
Finally, contrary to the expectation of Ref.\ \cite{RDV}, there is essentially
no correlation between the degree of bending present and the extension of
the scission configuration because the mode tends to equilibrate before
scission is approached.  However, the dynamical calculations do reveal an
effect on the spin-spin correlation: while the equilibrium value of $c_{AB}$
decreases from $\approx-0.08$ to $\approx-0.05$ when going from compact to
elongated scission configurations, the dynamical values decrease more,
namely from $\approx-0.08$ to $\approx-0.02$, but such small values
would probably be difficult to measure.

The present exploratory study considered a number of fission cases
at fairly high excitation energy where the microscopic modifications
of the potential energy are small and fast Langevin simulation is possible,
allowing the generation of large event samples.
To avoid complications from multi-chance fission,
the evaporation of neutrons during the shape evolution was turned off.
However, it would seem interesting to include pre-scission evaporation
in the calculation of the shape evolution because not only may the systems 
followed from saddle to scission be (slightly) different but, more importantly, 
they will typically be considerably colder 
which would reduce the spin fluctuations significantly

The focus has been on the nucleon-exchange mechanism
and contributions from other mechanisms, such as those considered in Refs.\ 
\cite{RandrupPRCL108,ScampsPRC108,ScampsPRC110,ShneidemanPRC111},
were not considered.
Therefore the present results are not meant to be specific predictions
but should rather be taken as indicative of the expected general behavior
of the correlated fragment spin distribution
generated by the nucleon-exchange mechanism.
In particular, for cases at lower excitation,
such as spontaneous or thermal fission, quantitative predictions
would require more elaborate calculations that include the microscopic
effects in the dynamics and employ suitably enlarged shape families.

\section*{Acknowledgments}
We thank R.\ Vogt for helpful discussions.
This work was supported in part by the Office of Nuclear Physics
in the U.S.\ Department of Energy's Office of Science 
under Contract No.\ DE-AC02-05CH11231
and by the Polish National Agency for Academic Exchange 
under the Ulam NAWA Programme (No\. BPN/ULM/2023/1/00113/DEC/1),
and the National Science Centre, Poland (grant No.\ 2023/49/B/ST2/00622).

\appendix
\section{Langevin simulation}
\label{Langevin}

In this study, the nuclear shape evolution during fission is simulated 
by a previously developed and extensively used Langevin model \cite{Adeev2005}
which is briefly described below.

The model simulates the time evolution of the nuclear shape
which is described by the Funny Hills parametrization \cite{BrackRMP44}
in terms of the elongation parameter $C$, the neck thickness $h$, 
and the mass asymmetry $\alpha$.
In cylindrical coordinates, the shape is given by
\beq
\rho_s(z)^2 = \left\{ \begin{array}{ll}
	 (C^2-z^2)(A_s+Bz^2/C^2+\alpha z/C), & B\geq0 ,\\
	 (C^2-z^2)(A_s+\alpha z/C)\exp(BCz^2) , & B<0 ,
\end{array}\right.
\eeq
where $z$ is the coordinate along the symmetry axis 
and $\rho_s$ is the transverse distance to the surface.
The parameters $B$ and $A_s$ are given by $B=2h+\half(C-1)$ and
\beq
A_s = \left\{ \begin{array}{ll}
	 C^{-3} + \mbox{$1\over5$}B , & B\geq 0 ,\\
-\mbox{$4\over3$}B/\left[e^x + (1+{1\over2x})
	\sqrt{-\pi x}\,{\rm erf}(-x)\right], & B<0 ,
\end{array}\right.
\eeq
with $x\equiv BC^3$.

In the stochastic approach \cite{AbePR275,PR292,NadtochyPRC72},
the evolution of the shape parameters is akin to Brownian motion with
the coupling to the microscopic degrees of freedom providing the random impulses.
The average effect of this coupling is a friction force,
while the fluctuating remainder, being modeled as Gaussian white noise,
gives the evolution a diffusive character.
Consequently, the shape evolution, and the resulting physical observables, 
will fluctuate.

Rather than working with $(C,h,\alpha)$ directly,
it is convenient \cite{Adeev2005,NadtochyPRC72}
to consider the shape coordinates $\bld{q}=\{q_i\}$ with $q_1=C$, 
$q_2=(h+\threehalf)/[5/2C^3+\quart(1-C)+\threehalf]$, 
and $q_3=\alpha(A_s+B)$ if $B\geq0$ or $q_3=\alpha/A_s$ if $B<0$.
The coupled Langevin equations of motion for these shape coordinates
and their conjugate momenta $\bld{p}=\{p_i\}$ are
\beqar
{dq_i\over dt} &=& \mu_{ij} p_j ,\\
{dp_i\over dt} &=& -\half p_ip_k{\del\mu_{jk}\over\del q_i}
-{\del F\over\del q_i} -\gamma_{ij}\mu_{jk}p_k +\theta_{ij}\xi_j ,\,\,
\eeqar
where summation over repeated indices is implied.
The random variables $\{\xi_i\}$ have normal distributions and the associated
random force $\bld{\theta}$ is related to the dissipation tensor $\bld{\gamma}$
by the Einstein relation, $\theta_{ik}\theta_{kj}=\gamma_{ij}T$,
where $T$ is the temperature of the heat bath, related to 
the internal nuclear excitation $E^*$ by the Fermi-gas relation, $E^*=aT^2$, 
where $a(\bld{q})$ is the level-density parameter.
Then $F(\bld{q}) = V(\bld{q})-a(\bld{q})T^2$ is the Helmholtz free energy,
where the potential energy $V(\bld{q})$ is calculated 
with a finite-range macroscopic model \cite{KrappePRC20}
using the parameters from Ref.\ \cite{SierkPRC33}.
Finally, $\bld{\mu}(\bld{q})$ is the inverse of the inertial mass tensor 
$\bld{m}(\bld{q})$ which is computed using the Werner-Wheeler approximation 
for incompressible irrotational flow \cite{DaviesPRC13}.
During the Langevin evolution,
the total energy is conserved, $E=V+E^*+K$,
where $K(\dot{\bld q},\bld{q})=\half \mu_{ij}p_ip_j$ 
is the collective kinetic  energy.

The dissipation tensor $\bld{\gamma}$ is based on 
the one-body dissipation mechanism \cite{onebody,RandrupAP125}.
For compact mononuclear shapes only the wall formula is used,
while the wall-plus-window formula is used for strongly necked-in shapes,
and a smooth interpolation is made for intermediate shapes.
In order to explore the role of the dissipation strength,
the present study varies the strength of the wall dissipation
by means of the parameter $k_s$.
The value $k_s=1$ gives the standard strength calculated theoretically
in the original work \cite{onebody}.
However, we have also used $k_s=0.25$ because a somewhat weaker dissipation, 
corresponding to $0.2\leq k_s\leq0.5$,
appears to be better suited for reproducing experimental data on
the mass-energy distribution and particle multiplicities \cite{Nix1986,Nix1987}.

Details about the potential energy surface, inertial-mass and friction tensors 
may be found in Refs.\ \cite{NadtochyCPC258,NadtochyCPC275}.

By solving the above Langevin equations, 
the dynamical evolution of the nuclear shape can be followed from
the ground state until scission.
It is then straightforward to extract the time dependence of
the quantities needed for the spin transport coefficients:
the distance between the mass centers, $R(t)$,
the neck radius, $c(t)$, the mass asymmetry, $\alpha(t)$,
as well as the temperature, $T(t)$.

\begin{figure}[t]	    
\includegraphics[trim={2.5cm 0.3cm 5.5cm 0cm},clip,width=0.25\textwidth,
     angle=-90]{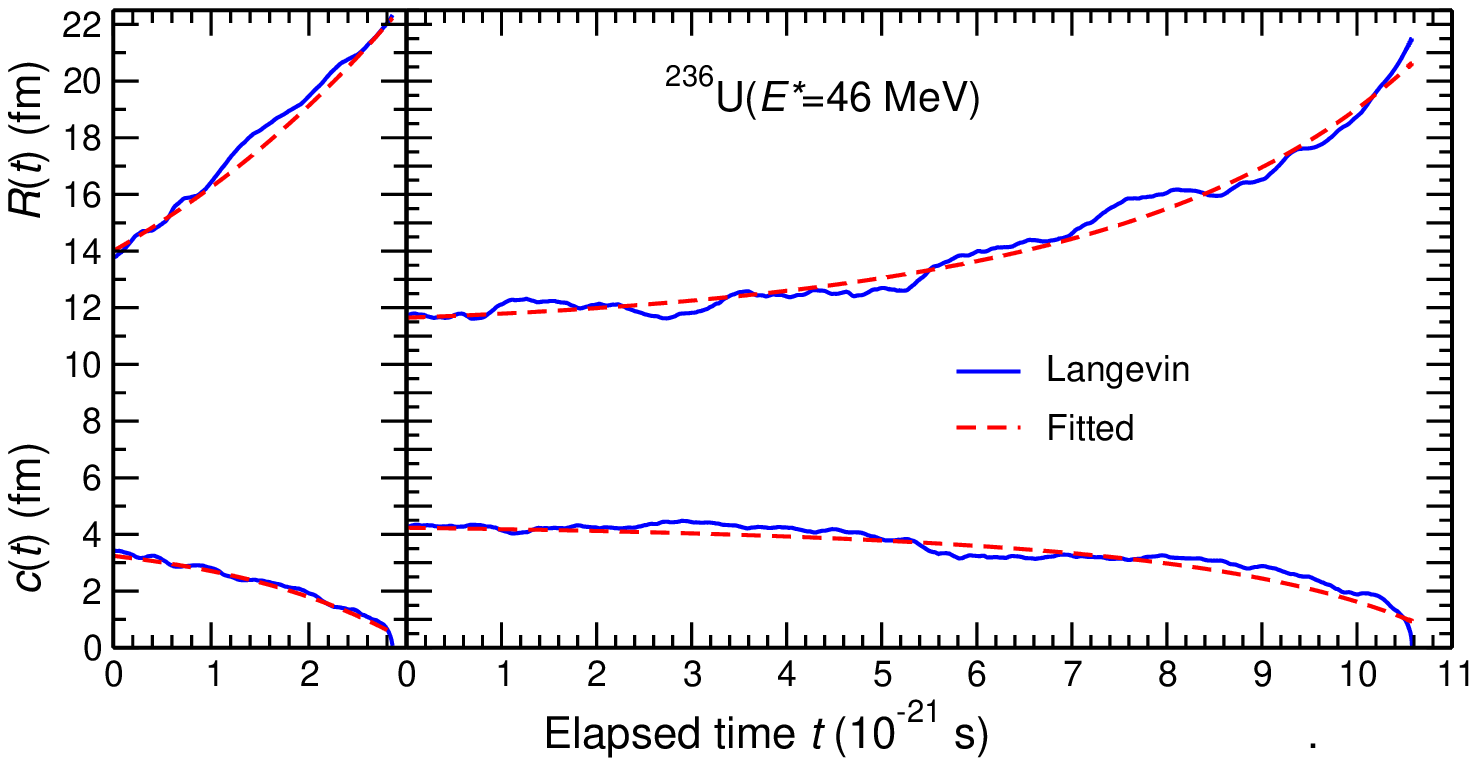}      
\caption{\label{f:Rc-fit}
The replacement of the Langevin result for $R(t)$ and $c(t)$ by the 
analytical form (\ref{fit}) for a fast and a slow event.}
\end{figure}	    	    

\begin{figure}[tbh]	    
\includegraphics[trim={2.5cm 0.3cm 5.5cm 0cm},clip,width=0.25\textwidth,
     angle=-90]{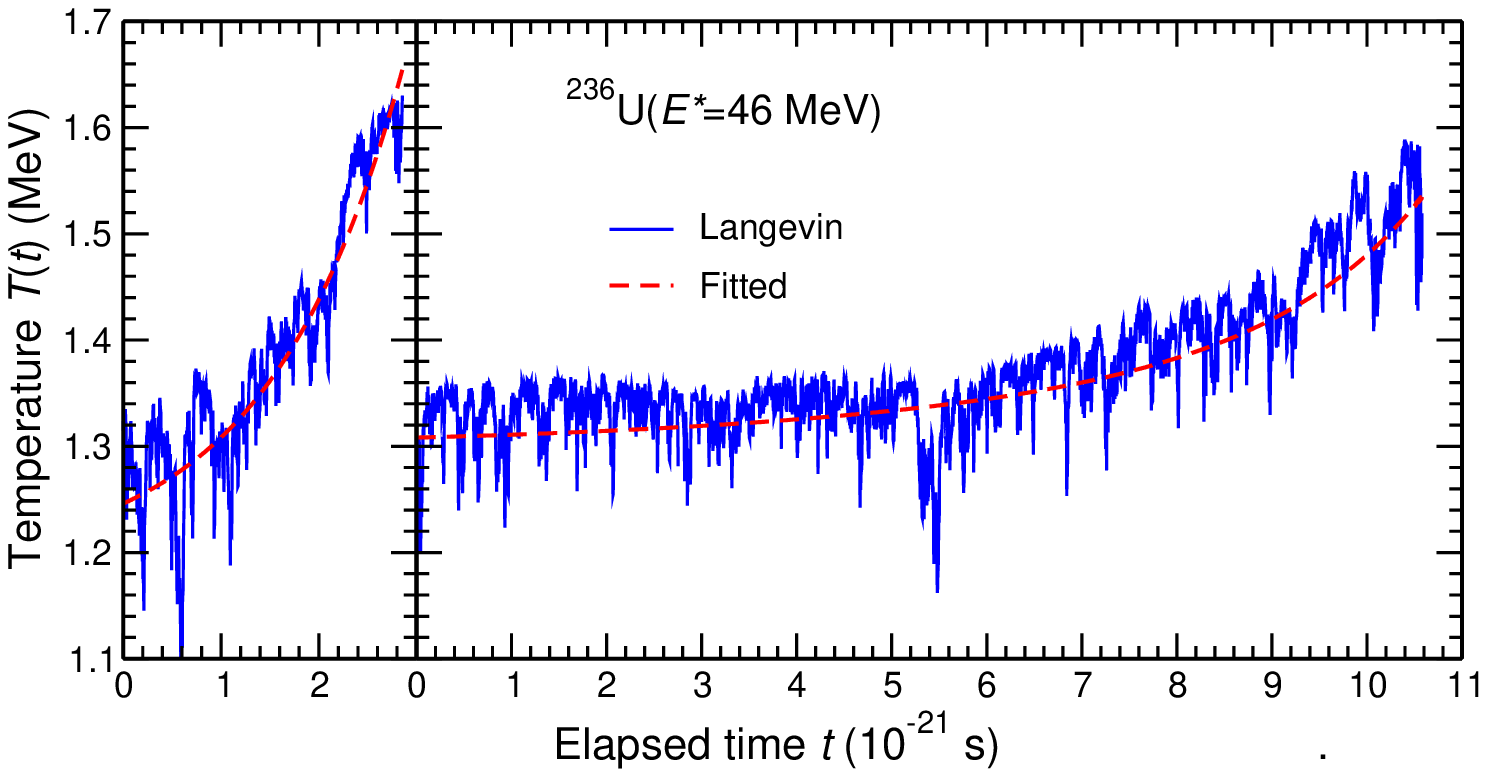}      
\caption{\label{f:T-fit}
The replacement of the Langevin result for $T(t)$ by the analytical form
(\ref{fit}) for the same events as in Fig.\ \ref{f:Rc-fit}.}
\end{figure}	    	    

As mentioned in the main text (Sect.\ \ref{shapes}),
the Langevin calculation employs time steps that are much
shorter than the time scale characteristic of the spin evolution and 
it is therefore convenient to replace the rapidly fluctuating Langevin results
for $R(t)$, $c(t)$, $\alpha(t)$, and $T(t)$ by smooth approximations.
Because of the difference in relevant time scales,
such a replacement has no impact on the calculated
evolution of the fragment spin distributions.

For each shape evolution, the Langevin result for the mass asymmetry parameter 
$\alpha(t)$ is replaced locally by linear fits, 
using typically twenty intervals.

Furthermore, the five-parameter form 
\beq\label{fit}
f(t) = c_1 + c_2\,e^{t/c_3} + c_4\,e^{(t/c_5)^2}
\eeq
is fitted globally to the Langevin result for $R(t)$, $c(t)$, or $T(t)$.
This simple smoothing is reasonably accurate,
as illustrated for a fast and a slow event
in Figs.\ \ref{f:Rc-fit} and \ref{f:T-fit}.



                        \end{document}